%% file: main.tex
\documentclass[letterpaper,twocolumn,10pt]{article}
\usepackage{usenix-2020-09} 

\input{preamble}

\begin{document}

\title{Empowering Data Centers for Next Generation 
Trusted Computing}
\date{\today} 		

\author{\rm{Aritra Dhar$^{\dag}$ \qquad Supraja Sridhara$^{\ddag}$ \qquad Shweta Shinde$^{\ddag}$ \qquad Srdjan Capkun$^{\ddag}$ \qquad Renzo Andri$^{\dag}$}  \\ \\ $^\dag${Computing Systems Lab, Huawei Zurich Research Center} ~\;~\;~\;~\;~\;~\;~\;~\;~\;~\; $^{\ddag}${ETH Z\"urich} \\ \\
}

\maketitle

\input{sections/abstract.tex}

\input{sections/introduction}

\input{sections/problem_statement}

\input{sections/design_space}

\input{sections/design}

\input{sections/case_studies_v3}

\input{sections/security_analysis_v2}

\input{sections/implementation_v3}

\input{sections/evaluation_v3}

\input{sections/related_work_v2}

\input{sections/conclusion}

\bibliographystyle{ieeetr}
\bibliography{bibliography} 

\appendix
\input{sections/appendix}

\end{document}

%% file: preamble.tex
\usepackage{color}
\usepackage{tabularray}
\UseTblrLibrary{siunitx}
\usepackage{array}

\usepackage{pifont}
\usepackage{xspace}
\usepackage{amssymb}
\usepackage{amsmath,amsthm}
\usepackage{booktabs,tabularx}
\usepackage{amsfonts}
\usepackage{enumitem}
\usepackage{xcolor,colortbl}
\usepackage[font=small]{caption}

\usepackage{listings}
\definecolor{mygreen}{rgb}{0,0.6,0}
\definecolor{mymauve}{rgb}{0.58,0,0.82}
\definecolor{mygray}{gray}{0.95}
\definecolor{mygray1}{gray}{0.6}

\usepackage{subfig}
\usepackage{xcolor}
\usepackage{graphicx}
\usepackage{pgf-umlsd}
\usepackage{tikz}
\usepackage{standalone}
\usepackage{multirow}
\usepackage{adjustbox}
\usepackage{wrapfig}

\usetikzlibrary{positioning}
\usetikzlibrary{shapes,arrows,calc,automata}
\usetikzlibrary{shapes.geometric}

\usepackage{hyperref}
\usepackage{url}

\usepackage{breakurl}
\usepackage{booktabs}

\hypersetup{
pdftitle={Distributed TEE Architecture},
pdfauthor={Aritra Dhar},
}

\usepackage{cleveref}
\Crefname{equation}{Eq.}{Eqs.}
\Crefname{figure}{Fig.}{Figs.}
\Crefname{tabular}{Tab.}{Tabs.}
\Crefname{section}{Sec.}{Sec
encircle.}

\usepackage{multirow}

\usepackage{cleveref}

\newif \ifSendETH
\SendETHfalse

\usepackage{graphicx}
\usepackage{tcolorbox}
\usepackage{array}
\newcolumntype{N}{@{}m{0pt}@{}}

\newcounter{observationCounter}
\crefname{observationCounter}{observation}{observations}
\Crefname{observationCounter}{Observation}{Observations}

\newcounter{requirementCounter}
\crefname{requirementCounter}{requirement}{requirements}
\Crefname{requirementCounter}{Requirement}{Requirements}

\newcounter{myctr}

\newenvironment{mylist}
    {\begin{list}{(\textbf{\arabic{myctr}})}
        {\usecounter{myctr}
        \setlength{\topsep}{1mm}\setlength{\itemsep}{0.5mm}
        \setlength{\parsep}{0.5mm}
        \setlength{\itemindent}{0mm}\setlength{\partopsep}{0mm}
        \setlength{\labelwidth}{-2mm}
        \setlength{\leftmargin}{1mm}}
    }
    {\end{list}}

\newenvironment{mybullet}
    {\begin{list}{$\bullet$}
        {\setlength{\topsep}{1mm}\setlength{\itemsep}{0.5mm}
        \setlength{\parsep}{0.5mm}
        \setlength{\itemindent}{0mm}\setlength{\partopsep}{0mm}
        \setlength{\labelwidth}{-2mm}
        \setlength{\leftmargin}{1mm}}
    }
    {\end{list}}

\newcommand{\myparagraph}[1]{\noindent\textbf{#1.}}

\newcommand{\spacesave}{\vspace{0pt}}

\let\oldding\ding
\renewcommand{\ding}[2][1]{\scalebox{#1}{\oldding{#2}}}

\newcommand{\fdu}{FDU\xspace}
\newcommand{\fdus}{FDUs\xspace}

\newcommand{\ertfull}{Enclave Routing Table\xspace}
\newcommand{\ert}{ERT\xspace}

\definecolor{mygreen1}{RGB}{169, 209, 142}
\definecolor{myyellow1}{RGB}{255, 230, 153}
\definecolor{myblue1}{RGB}{180, 199, 231}
\newcommand{\green}[1]{\colorbox{mygreen1}{#1}}

\newcommand{\splitatcommas}[1]{%
  \begingroup
  \begingroup\lccode`~=`, \lowercase{\endgroup
    \edef~{\mathchar\the\mathcode`, \penalty0 \noexpand\hspace{0pt plus 1em}}%
  }\mathcode`,="8000 #1%
  \endgroup
}

\newcolumntype{?}{!{\vrule width 1pt}}

\theoremstyle{definition}

\newcommand{\q}[1]{\textcolor{blue}{#1 ?}}

\newcommand{\manifestid}{$\tt{manifestID}$\xspace}

\newcommand{\user}{user\xspace}
\newcommand{\verifier}{verifier\xspace}

\newcommand{\job}{job\xspace}

\newcommand{\fmt}{FMT\xspace}
\newcommand{\acu}{ACU\xspace}
\newcommand{\mpe}{MPE\xspace}
\newcommand{\acuAICore}{ACU$_{core}$\xspace}
\newcommand{\acuAIMem}{ACU$_{mem}$\xspace}

\newcommand{\loc}{LoC\xspace}

\newcolumntype{R}[2]{%
    >{\adjustbox{angle=#1,lap=\width-(#2)}\bgroup}%
    l%
    <{\egroup}%
}

\definecolor{Gray}{gray}{0.85}

\newcolumntype{a}{>{\columncolor{Gray}}c}
\newcolumntype{b}{>{\columncolor{white}}c}

\colorlet{punct}{red!60!black}
\definecolor{background}{gray}{0.99}
\definecolor{delim}{RGB}{20,105,176}
\colorlet{numb}{magenta!60!black}
\definecolor{eclipseStrings}{RGB}{42,0.0,255}
\definecolor{eclipseKeywords}{RGB}{127,0,85}
\definecolor{codegreen}{rgb}{0,0.6,0}

\lstdefinelanguage{json}{
    basicstyle=\bfseries\scriptsize\ttfamily,
    showstringspaces=false,
    breaklines=true,
    frame=lines,
    backgroundcolor=\color{background},
    morekeywords={TRUE,FALSE,linkEnc},
    keywordstyle=\color{numb},
    literate=
     *{0}{{{\color{numb}0}}}{1}
      {1}{{{\color{numb}1}}}{1}
      {2}{{{\color{numb}2}}}{1}
      {3}{{{\color{numb}3}}}{1}
      {4}{{{\color{numb}4}}}{1}
      {5}{{{\color{numb}5}}}{1}
      {6}{{{\color{numb}6}}}{1}
      {7}{{{\color{numb}7}}}{1}
      {8}{{{\color{numb}8}}}{1}
      {9}{{{\color{numb}9}}}{1}
      {:}{{{\color{punct}{:}}}}{1}
      {,}{{{\color{punct}{,}}}}{1}
      {\{}{{{\color{delim}{\{}}}}{1}
      {\}}{{{\color{delim}{\}}}}}{1}
      {[}{{{\color{delim}{[}}}}{1}
      {]}{{{\color{delim}{]}}}}{1},
}

\lstdefinelanguage{ccode}{
    basicstyle=\bfseries\scriptsize\ttfamily,
    showstringspaces=false,
    commentstyle=\color{codegreen},
    language=C,
    breaklines=true,
    frame=lines,
    morecomment=[f][\color{green}][0]{*},
    morecomment=[f][\color{red}][0]{\#},
    backgroundcolor=\color{background},
    morekeywords={TRUE,false},
    keywordstyle=\color{numb},
    literate=
     *{0}{{{\color{numb}0}}}{1}
      {1}{{{\color{numb}1}}}{1}
      {2}{{{\color{numb}2}}}{1}
      {3}{{{\color{numb}3}}}{1}
      {4}{{{\color{numb}4}}}{1}
      {5}{{{\color{numb}5}}}{1}
      {6}{{{\color{numb}6}}}{1}
      {7}{{{\color{numb}7}}}{1}
      {8}{{{\color{numb}8}}}{1}
      {9}{{{\color{numb}9}}}{1}
      {:}{{{\color{punct}{:}}}}{1}
      {,}{{{\color{punct}{,}}}}{1}
      {\{}{{{\color{delim}{\{}}}}{1}
      {\}}{{{\color{delim}{\}}}}}{1}
      {[}{{{\color{delim}{[}}}}{1}
      {]}{{{\color{delim}{]}}}}{1},
}

\usepackage{subfig}

%% file: sections/abstract.tex
\begin{abstract}

Modern data centers have grown beyond CPU nodes to provide domain-specific accelerators such as GPUs and FPGAs to their customers. From a security standpoint, cloud customers want to protect their data.
They are willing to pay additional costs for trusted execution environments such as enclaves provided by Intel SGX and AMD SEV. Unfortunately, the customers have to make a critical choice---either use domain-specific accelerators for speed or use CPU-based confidential computing solutions. 

To bridge this gap, we aim to enable data-center scale confidential computing that expands across CPUs and accelerators. We argue that having wide-scale TEE-support for accelerators presents a technically easier solution, but is far away from being a reality. Instead, our hybrid design provides enclaved execution guarantees for computation distributed over multiple CPU nodes and devices with/without TEE support. Our solution scales gracefully in two dimensions---it can handle a large number of heterogeneous nodes and it can accommodate  TEE-enabled devices as and when they are available in the future. We observe marginal overheads of $0.42$--$8\%$ on real-world AI data center workloads that are independent of the number of nodes in the data center. We add custom TEE support to two accelerators (AI and storage) and integrate it into our solution, thus demonstrating that it can cater to future TEE devices.
\end{abstract}

%% file: sections/introduction.tex
\section{Introduction}
\label{sec:introduction}

The cloud service providers (CSP) have moved away from CPU-centric monolithic design where CPU, memory, and devices are on the same physical platform~\cite{disaggregatedcomp}.
Instead, domain-specific accelerators or DSAs have become an essential part of modern data centers~\cite{a100, yesil2015hardware, ham2016graphicionado} for workloads such as AI/ML~\cite{keras, choi2022serving}, graph processing~\cite{zhu2016gemini, gonzalez2014graphx}, in-memory cache~\cite{lavasani2013fpga, chalamalasetti2013fpga}, and many more. 
DSAs are often deployed as standalone nodes in a rack with network functions~\cite{eran2019nica} to communicate with other nodes~\cite{lu2018multi, singhvi20201rma}. 
DSAs are shared between multiple tenants~\cite{aljahdali2014multi} concurrently to maximize resource utilization and scalability.

The heterogeneity in the modern multi-tenant cloud poses new security challenges for ensuring the integrity and confidentiality of users' data and execution. 
Specifically, the attack surface includes mutually distrusting co-tenants as well as the CSP itself. 
Reasoning about an untrusted CSP requires protecting against a large attack surface, including all the CSP software and intermediate devices necessary for running the cloud. 
More importantly, it necessitates stopping the CSP from maliciously reconfiguring the nodes being used by tenants~\cite{adler2011leap}.

Trusted execution environments (TEEs) are one of the building blocks to enable a secure multi-tenant cloud. 
They provide \textit{enclaves} to execute user code without trusting the OS or hypervisors~\cite{costan2016intel, winter2008trusted, costan2016sanctum, suzaki2011memory}.
CPUs~\cite{sgx, TZOS, keystone, TDX, SEVVM, CCA} as well as on DSAs~\cite{volos2018graviton, h100, kang2021iceclave, zhao2021shef, zeitouni2021trusted} support hardware-enabled enclaves.
Given their effectiveness, TEEs have become an important offering in the cloud~\cite{azure_confidential_cloud, google_confidential_computing,ibm_confidential_cloud, AWS_nitro_enclave}.
Although TEEs serve a necessary role, they are not sufficient if the user wants to fully use the heterogeneity in a modern multi-tenant cloud. 

\begin{figure}[]
  \centering
    \includegraphics[trim={0 13.6cm 26cm 0},clip,width=0.75\linewidth]{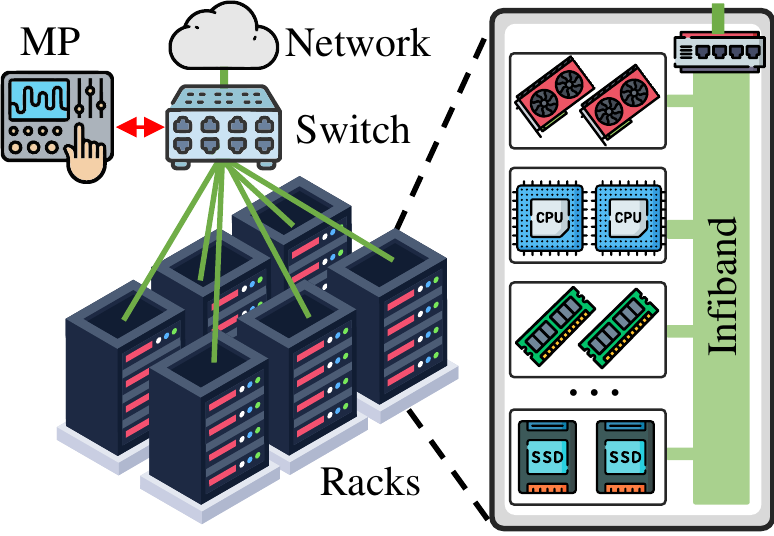}
    \caption{\textbf{Data center topology.} Racks are connected over high-speed optical switches. Inside a rack, multiple nodes of the same type are
     clustered together and then connected over InfiniBand which communicates with the data center-level switch. The management plane (MP) performs resource allocation, revocation, and
     monitoring. }
    \label{fig:datacenter_topology}
\end{figure}

Only a small fraction of the data-center nodes that can be rented out to tenants are TEE-enabled.
Consider a tenant who wants to perform a computation that is best suited for an AI accelerator node. With state-of-the-art solutions, such a tenant has to choose between performance and security. 
Either execute their computation on a TEE-enabled node (e.g., CPU) which is not the optimal device, thus sacrificing performance. Or execute on the non-TEE AI accelerator in an unprotected environment. 
One potential solution to address this problem is to make all nodes TEE-enabled, but this is not a practical. 
For protecting non-TEE nodes, prior works have shown the feasibility of doing so via bus-level isolation if they are directly physically connected to a TEE host~\cite{schneider2022composite, bahmani2020cure, cronus}.
Such host-centric solutions do not apply to a data-center setting where nodes are connected in clusters and racks as shown in~\Cref{fig:datacenter_topology}.
On the other hand, solutions that enable TEE abstractions for a single rack containing non-TEE nodes seem promising~\cite{zhu2020hetee}. 
However, they do not scale to multiple racks and are not designed to leverage nodes that have TEE support. 
Lastly, simply connecting TEE nodes and rack-level solutions are not sufficient, it does not make them collectively secure. 
Doing so requires careful design considerations for security as well as scalability at a data-center level.

In this paper, we investigate the feasibility of \textit{enclaved} execution
that encompasses multi-tenant heterogeneous nodes that go beyond just
TEE-enabled CPUs. 
We design a distributed TEE solution that allows a tenant to securely
use TEE nodes (including CPUs and DSAs) and non-TEE legacy nodes. 
We use three main insights to achieve this goal. 
First, we use TEEs on CPUs and DSAs when available. Further, we use
such TEEs to protect all the data leaving the corresponding nodes. 
Second, we employ a centralized security controller (SC) that shields all
the non-TEE nodes. All non-TEE nodes are placed behind our trusted
controller who imparts TEE properties such as attestation, isolation,
and secure channel.

Put together, these two design decisions seem trivial and sufficient
to ensure that execution on each node is protected. 
But, on a closer look, it leaves several gaps that need to be
addressed. 
For example, the CSP can reconfigure nodes such that multiple tenants
have overlapping resources or turn off protection mechanisms while
the tenant is executing sensitive computations. 
A physical attacker can bypass our security controller, say by
directly connecting non-TEE nodes physically.
Put in other words, local protection at each node is not sufficient,
we need to ensure global isolation.
Our third insight addresses these gaps systematically.
We start by ensuring that the initial state of the nodes is attested
and cannot be changed thereafter. 
The controller checks that the CSP's resource management decisions do
not violate resource isolation and secure path guarantees. 
We provide a secure physical perimeter to a collection of non-TEE
nodes to make them equivalent to the TEE node's protection against a
physical adversary. 
In summary, we approach the problem in a distributed fashion by first
ensuring local security and then enforcing global properties to
maintain security.

We demonstrate our proposed solution by prototyping its main 
components.
We showcase the hardware and software changes required to add TEE support to DSAs. Specifically, we build hardware TEE primitives for
a state-of-the-art AI accelerator and several commercially available SSDs.
We synthesize our hardware design to show that they are feasible with a reasonable area and power cost.
We then demonstrate that our design is compatible with existing TEE-enabled GPUs and FPGAs.
We build a prototype for our controller to show the feasibility of our
end-to-end design for TEE and non-TEE nodes. 
We evaluate our system on real-world data center workloads. Our
cycle-accurate simulation shows that the scalability, performance,
and compatibility cost of our solution are within the reach of
practical deployment. 

On average, our accelerator incurs an overhead of 12\% during setup, 1.0141\% during inference runtime, and $4.4$ ms of teardown time.
Our synthesized hardware design only consumes 0.38 $\mu$W ($1.15\times 10^{-7}\%$) 2786 $\mu m^2$ ($0.2\%$) area on the accelerator die. 
The modification on the SSDs only incurs an average 16\% \& 20\% overhead for random sequential read-write benchmarks. Integrating existing GPU and FPGA only incurs $1.89\%$ and $0.04\%$ overhead respectively.
Our design scales with concurrent multi-tenants and the SC can support 912 concurrent nodes running inference workload in full bandwidth. In summary, we make the following contributions:
\begin{mybullet}
    \item {\bf \em Design.} We explore the design space for enclave
     execution in a heterogeneous setting of a multi-tenant cloud
     environment. We show the feasibility of using TEE and non-TEE
     nodes to provide enclave security primitives.

    \item {\bf \em TEEs on DSAs.} We show the hardware changes
     necessary for multi-tenant enclaved execution on DSAs by
     building it ourselves (AI accelerator and an SSD) and by
     re-using prior proposals (GPUs and FPGAs).

    \item {\bf \em Evaluation.} We provide simulation results with
     real-world workloads that show the scalability 
     and modest performance costs of our solution.
\end{mybullet}

%% file: sections/problem_statement.tex
\begin{figure*}[t]
  \centering
    \includegraphics[trim={0cm 13.5cm 3cm 0}, width=0.85\textwidth]{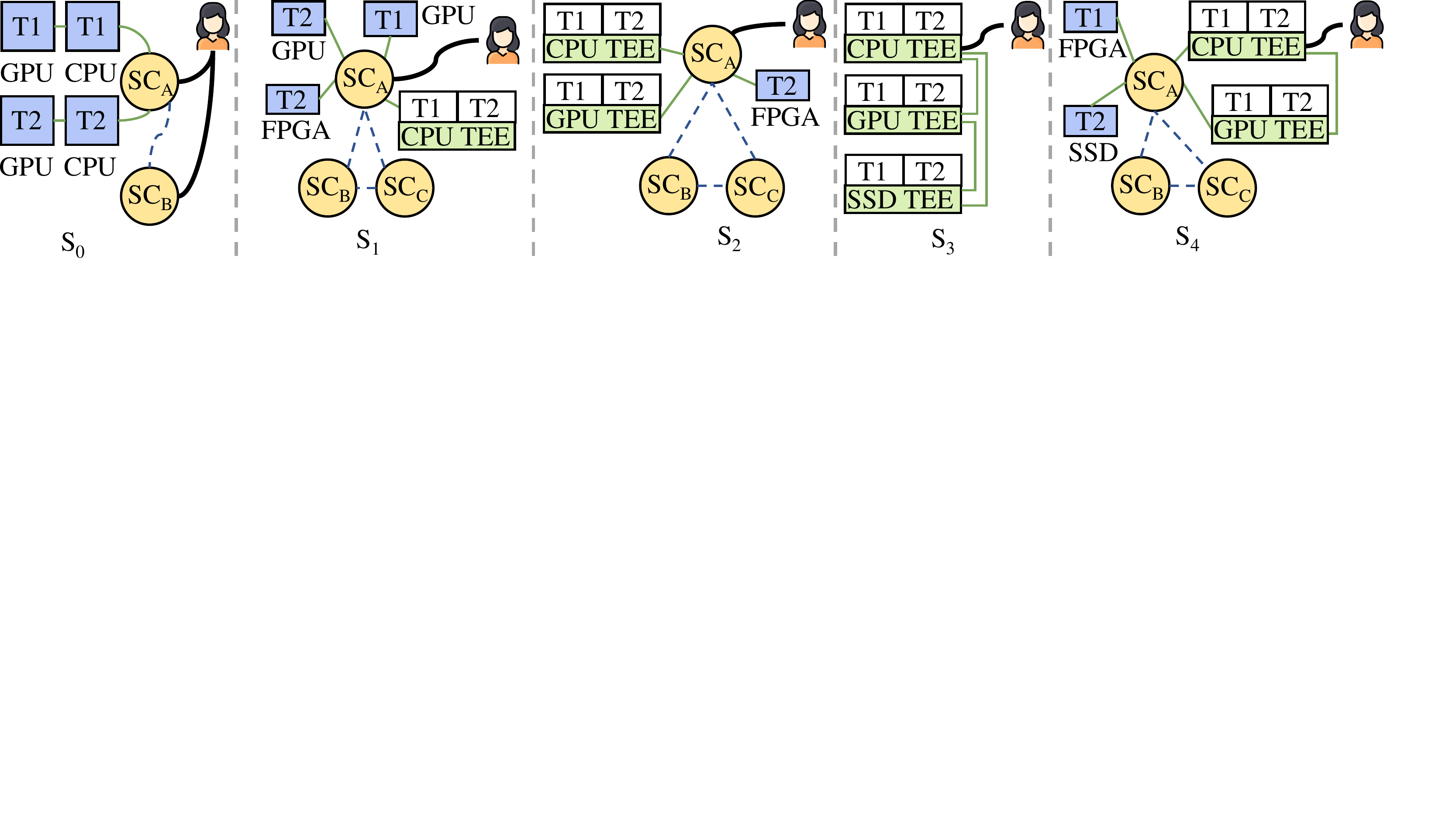}
    \caption{\textbf{Data center-scale heterogeneous TEE designs}. The figure shows five potential design choices: $S_{0-4}$. Nodes assigned to tenants T1/T2}
    \spacesave
    \label{fig:design_decisions}
\end{figure*}

\section{Motivation}
\label{sec:problemStatement}

\textbf{\myparagraph{Setting \& Problem Statement}}
Modern data centers enable \emph{disaggregated} execution (e.g., dReDBox project~\cite{dis1,dis2}) to improve the performance of computation by using domain-specific accelerators (DSA) such as GPUs, ML/AI accelerators, storage accelerators~\cite{legtchenko2017understanding}, and smart NICs~\cite{fungible_disagg}. \Cref{fig:datacenter_topology} shows a modern datacenter topology consisting of racks connected over a fast interconnect. 
Each rack has several CPU or DSA nodes where the CPU nodes are TEE-capable and a DSA node could either be a TEE or a non-TEE (legacy) node.
The resources, e.g., core, cache, memory, etc. on the TEE nodes can be divided dynamically into smaller units based on demand. 
We call these units: \textit{fungible device units} (\fdu) and they are functionally identical to the underlying nodes.
We assume the nodes (CPUs, GPUs, FPGAs, AI accelerators) have \textit{virtualization} layers (hypervisor).
Therefore both the non-TEE nodes and \fdus on the TEE nodes can be shared between multiple tenants. The CSP uses a management plane for resource allocation, revocation, and monitoring~\cite{lim2009disaggregated,lin2020disaggregated,tsai2020disaggregating}. 
A tenant deploys a \job by submitting a manifest to the management plane. The manifest specifies the size and type of resources the \job needs. In this disaggregated setting, a \job can run on multiple different types of \fdus. 
We ensure that trusted \fdus are provisioned on both the CPUs and devices as specified in the manifest. 

We investigate the design principles that are building blocks for enabling trusted computing in multi-tenant heterogeneous data centers. Specifically, we want to ensure that the programs and the data from the user stay protected even when the CSP, their infrastructure, and software stack are attacker-controlled. 

\myparagraph{Threat Model} We assume enclave code (CPU enclave, device drivers) running on CPU cores and \fdus (DSA code) are trusted. 
Therefore, a tenant trusts all code that they run on all their \fdus for a \job. 
However, co-located enclaves from different tenants are mutually distrusting. 
The CSP can deploy non-TEE and TEE nodes in the data-center. 
For TEE nodes, we trust the hardware manufacturer and these nodes have hardware root-of-trust for generating correct attestation reports and securing keys. 
We assume that an up-to-date revocation list is available to determine if an attestation report is from a revoked node.
Additionally, we trust the node's TEEs software TCB (e.g., the trusted hypervisor in ARM CCA, trusted security monitor in Keystone) including the firmware after verifying its integrity.  

The attacker can control the host OS's, hypervisors, the management plane, and other applications running on co-located \fdus. The attacker also controls all nodes and can plug in and out malicious nodes and network devices (e.g., switches, routers, network interconnects, etc.). The chipset and BIOS on the motherboards are also attacker-controlled, making PCIe link encryption and attestation mechanisms~\cite{pcie_attest} untrustworthy.
We assume a physical attacker who can probe the bus and manipulate all the network traffic. 

Finally, we leave Iago~\cite{checkoway2013iago}, side-channels, hardware trojans, and denial-of-service out of scope for this paper.

%% file: sections/design_space.tex
\section{Distributed TEE Design Space Exploration}
\label{sec:design_space}

\subsection{Potential Designs}
\label{sec:design_space:potential_designs}

We explore five potential ways ($S_{0-4}$) to design a data center-scale
trusted computing infrastructure, as shown in ~\Cref{fig:design_decisions}.

\myparagraph{$S_0$: Centralized SC between non-TEE CPUs} 
We assume that none of the nodes, CPUs or DSAs, are TEE enabled. Each
DSA is physically connected to a CPU node and they both are allocated
entirely to a single tenant. 
In such a topology, any tenant can tamper any node's memory or the
communication channels. 
We employ a simple solution where all CPUs are physically connected to
a central trusted security controller (SC).
The SC is responsible for resource isolation and does so by monitoring
all communication between CPUs as well as with remote users. 
For local access, the SC ensures that nodes of the same tenant can
access each other. To protect the communication channel between the
tenant and the nodes, the SC provides a secure channel.
Such an SC can be implemented in different ways and can be a dedicated
hardware module~\cite{zhu2020hetee}. 
This design does not allow multi-tenant devices as nodes are
assigned exclusively to a single tenant. As an SC can be physically
connected to a limited number of nodes, scaling the design beyond a
single rack requires reasoning about global inter-SC co-ordination in
the presence of untrusted nodes. Lastly, the SC has to be aware of
all allocation decisions in order to enforce isolation.

\myparagraph{$S_1$: Centralized SC between CPU TEEs and DSA nodes}
We assume that the CPU nodes have TEE and can be virtualized. 
However, the DSA nodes do not have TEE support. 
In such a topology, we choose to connect the DSA nodes directly to the
SC, instead of being connected to the CPU as in $S_0$.
$S_1$ scales better because DSAs can communicate directly,
without going via CPUs. Each unit on a virtualized CPU can be
assigned to dedicated DSAs, thus improving multi-tenancy.
To ensure isolation, the SC intercepts all the traffic from the
user and between the nodes and performs access control such that
tenants cannot access each other's devices.
If the tenant communicates directly with a DSA, the SC provides a
secure channel to ensure that the incoming and outgoing data is
protected. 
However, this makes the SC a bottleneck without improving the scale
beyond a single rack---inter-SC protection still requires a global
view while being aware of all allocation decisions.

\myparagraph{$S_2$: Semi-centralized SC with TEE and non-TEE nodes}
TEE support goes beyond just CPU nodes as shown in recent proposals
for peripheral TEEs~\cite{volos2018graviton, zhao2021shef, kang2021iceclave, h100}.
For nodes that are TEE-enabled, we can support the secure isolation of
tenants without requiring the SC.
Specifically, we can shift the SC's access control checks to local
monitors on the TEEs. 
However, all non-TEE node communications have to always go through the SC. 
Although better than $S_1$ and $S_2$, the SC still needs a global
resource allocation view and handles communication with the user and
non-TEE nodes.

\myparagraph{$S_3$: Decentralized TEEs} 
We can eliminate the SC by requiring that all nodes are TEE-enabled. 
Such nodes enable fully decentralized TEE, where they can simply set
up secure channels between the nodes after attestation~\cite{mage, opera}. Further, the
user can directly communicate with the node, therefore, eliminating
any potential bottlenecks. 
Such a decentralized design allows the TEE nodes to be securely
virtualized and be allocated to tenants without the need for a
trusted SC, making it suitable for modern multi-tenant data
centers. 
However, it has a strong requirement that all nodes have hardware TEE
support, making it infeasible for many devices including legacy
DSAs.

\myparagraph{$S_4$: Hybrid} 
We need a design that accommodates upcoming TEE-capable DSAs, existing
TEEs, legacy DSAs, and devices without hardware TEE capability. 
Noting from $S_3$, TEE-capable devices can be completely decentralized
without the need for an SC. 
Therefore, such nodes can leverage virtualization and scale across
many racks. 
Noting from $S_1$, we can use the SC to protect non-TEE devices. 
We can carefully allocate non-TEE devices, such that they are all
behind an SC via a physical connection. 
We can use the SC to mediate all communication between non-TEE nodes for access control, in the same fashion as in $S_1$.

Thus, SC acts as the security monitor that provides a secure channel and
filters the traffic to the non-TEE nodes based on the access
policies. In other words, it performs the TEE protection on the
behalf of the non-TEE nodes. 
Lastly, the SCs across multiple racks only have to keep track of local non-TEE nodes. 
Thus, such a design provides the most security and functionality
suitable for multi-tenant data centers. 

\subsection{Security Considerations}
\label{sec:design_space_observations}

Although intuitively $S_4$ seems to be the optimal design choice,
reasoning about its security requires careful consideration.


\myparagraph{Challenges}
While monitoring data on the bus and secure communication between nodes is necessary it is insufficient to determine the local node's configuration state. 
First, modern DSAs are highly configurable, i.e., the CSP or
hypervisor can push configuration to the DSAs' firmware. For example, 
hypervisors allow pass-through access to peripherals, and can reconfigure their partition size ~\cite{adler2011leap}. 
NVIDIA A100 GPUs~\cite{mig} allow their GPU partitions (known as multi-instance GPU or MIG) to reconfigure partition size. 
Second, even if we ensure that the firmware and configuration of the
device are correct after bootup, an attacker-controlled
OS/hypervisors can reconfigure a device or flash its firmware later.
So, it is crucial to ensure that the device initialized from a safe
state stays in a safe state. This is necessary for both TEE and
non-TEE-capable devices. Thus, the user or the SC needs to ensure
that the integrity of the device firmware is maintained before and
after secret provisioning.
Third, the SC is not aware of the execution of the DSAs due to their asynchronous nature. 
Therefore, misbehaving programs on rogue devices can attempt to bypass the SC and directly access other tenants' data. 
Thus, it is necessary to ensure that all communications between nodes are mediated by either the TEEs or the SC.
Finally, we have to protect the SC, TEE, and non-TEE nodes, and all the communication channels from a physical adversary.

\myparagraph{Insights}
We can use secure/measured boot to ensure that both the TEE and non-TEE-capable devices' firmware and configuration are correct. 
The SC must check that integrity of the all device firmware is
maintained and should relay this to the user before provisioning secrets.
We can use attestation to ensure that the DSAs are running the
user-provided code (also known as the device kernels).
TEE-level isolation ensures that the attacker cannot access the memory
of other tenants on the same node. For non-TEE devices, the SC knows
the resource allocation per tenant. At runtime, it ensures tenant
isolation by performing bus-level access control which checks if the
accessor is authorized to access the resource. 
The communication channel between the nodes is protected, i.e., 
an attacker cannot manipulate the data on the bus.
All DSAs require secure and persistent secure on-device key storage to
execute remote attestation and secure boot. 

\myparagraph{Physical Attacker} We assume that CPU-TEE nodes have hardware protection such as MEE in SGX, to defend against a physical attacker. 
We assume that that the memory on a TEE-DSA node (e.g., Graviton~\cite{volos2018graviton}, SheF~\cite{zhao2021shef}), 
is part of the chip that an attacker cannot infiltrate, such as---high performance 3D stacked memories~\cite{woo2010optimized} for accelerators~\cite{amd_3d_memory}, GPU~\cite{nv_3d_memory}, and SoCs~\cite{beyne20213d}.
For off-core memory, memory encryption and integrity protection proposals on GPU~\cite{na2021common,yuan2021analyzing}, NPU~\cite{lee2022tnpu,hua2022guardnn}, and data center accelerators~\cite{10.1145/3470496.3527418} can thwart physical attackers with some performance penalties.
However, protecting non-TEE nodes is a significant challenge as no such protection mechanisms are available for legacy nodes.
We assume that the SC and TEE nodes are secured in a physical container that can detect and prevent tampering~\cite{nisarga2016system,rack_level_security}. Such mechanisms are used in existing proposals~\cite{zhu2020hetee}. Other communications channels between TEE nodes, TEE nodes \& SC, and between SCs are integrity and confidentiality protected.

\myparagraph{TCB} The TCB includes the following: hardware components such as the CPU, and DSA-SoCs are trusted. The SC software and hardware, and device firmware are trusted.

\subsection{Design Requirements}
\label{sec:design_space:overview}

Our solution builds on top of existing data center architecture by keeping the cloud service provider's management plane (CSP-MP) unchanged.
We assume that MP, software stack (OS, hypervisors, other applications), and data center infrastructure (NIC, switches, interfaces) are attacker-controlled. 
The CSP-MP is responsible for resource allocation/deallocation, load balancing, workload deployment, scaling, and management.
All nodes have a hardware root of trust (RoT) that can be used for securely
bootstrapping the node. For TEE-capable devices, this RoT
is used for remote attestation to ensure the integrity of the
firmware and the \fdus before the remote users provision secrets. On the TEE-capable nodes, the security monitor or
SM, a high-privileged trusted entity on the device, provides attestation and enforces isolation between \fdus.
The SM can split the nodes into several multi-tenant units (\fdu). 

For non-TEE devices, we need external trusted hardware, such as a
security controller (SC), in the rack. The SC is a low-TCB hardened
trusted entity that can be verified and mediates the trusted path
from a tenant to a device.
Users provide a manifest that describes the resource requirements, as shown in~\Cref{fig:manifest} in~\Cref{sec:appendix:manifest}.
Note that our approach assumes that the programmer provides a static maximum memory size required for each \fdu.
Once the resources (TEE and non-TEE) are allocated and the
hypervisors are made aware of it, we have to identify and set up MMIO
as well as DMA regions and isolate them. 
Further, to ensure that all MMIO and DMA transactions are
confidentiality and integrity protected between nodes, we have to
establish a secure channel between \fdus.

%% file: sections/design.tex
\section{Design}
\label{sec:system}

Our design secures both non-TEE and TEE-capable nodes. 
We ensure that a CPU node is \textit{always TEE-enabled} as it runs a attacker-controlled OS or hypervisor. Additionally, our design requires that the user allocates at least one CPU-TEE node. 
We \textit{do not allow} a non-TEE node to be multi-tenant. 
We assume that memory allocation is \textit{static} and can be decided when an enclave is set up. We observe in a majority of DSAs, the size of the programs (known as the \textit{kernels}) and their memory allocations are predefined in the programs. The DSAs' programming model does not allow new memory allocation (e.g., \texttt{malloc}) while inside the kernel. For other devices such as storage or memory, the user can define the size based on the workload requirements. 
Therefore, a programmer can statically program the physical memory sizes of each remote \fdu into the enclave. 

\begin{figure}
  \centering
    \includegraphics[trim={0 15.22cm 22cm 0},clip,width=0.95\linewidth]{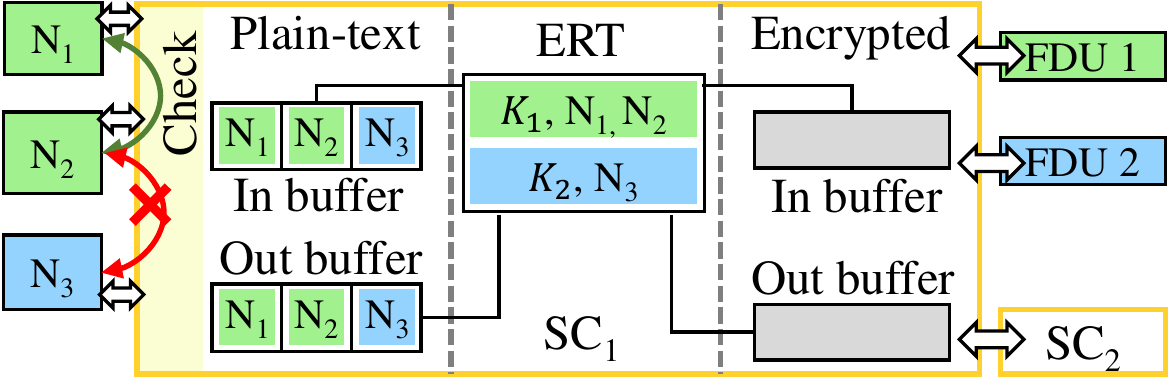}
    \caption{\textbf{Security Controller (SC)} enforces access control between TEE \& non-TEE nodes, and within non-TEE nodes.}
    \spacesave
    \label{fig:sc}
\end{figure}

\myparagraph{Setup} The user submits the number, size, and type of each \fdu and non-TEE node in a manifest file (example in~\Cref{sec:appendix:manifest}) along with the code to be executed on each \fdu and non-TEE node to the CSP-MP. 
The CSP-MP allocates the requested resources and collect attestation reports from them. 
The user gets the attestation reports and verifies them. Upon successful verification, the user sets up a secure channels with all the resources and provisions the job-specific secret key ($\mathcal{K}$) to start the computation. 
We now discuss how to secure both non-TEE and TEE nodes, and how to compose a rack with both types of nodes.

\subsection{Securing Non-TEE Nodes with SC}
\label{sec:system:non-tee}

\myparagraph{Security Controller (SC)} It is a device with hardware root-of-trust that can perform memory isolation for itself. 
The SC acts as the \textit{TEE proxy} for the non-TEE nodes and provides TEE properties such as isolation and trusted path for them. 
SC ensures that a non-TEE node is allocated to a single tenant and is not being shared with any other tenant or entity (e.g., CSP).
SC protects all communications between TEE and non-TEE memory by performing authenticated encryption and decryption using the shared secret ($\mathcal{K}$) provisioned during user setup. 
For communication between non-TEE nodes the SC does not perform authenticated encryption; instead it enforces access control. 
Specifically, it checks if source and destination for a transaction belong to the same job and if so allows it; otherwise it blocks it. 

The SC is connected physically (over PCIe) to the non-TEE nodes. 
During start-up, the SC first \textit{resets} the non-TEE nodes to its factory configuration and creates locally unique identities (e.g., PCIe physical ids which cant be forged) for them.
The SC maintains input and output buffers for the non-TEE nodes. Since the data in these buffers are in plain text, the SC needs to enforce isolated access to them. 
Specifically, it allocates non-overlapping regions of the buffers to individual non-TEE nodes as shown in~\Cref{fig:sc}. This ensures that non-TEE nodes cannot access each other's plain text data. 
To enforce access control between non-TEE nodes, the SC maintains a special data structure called an {\em \ertfull (\ert)}.
The \ert keeps track of the nodes that are a part of the same job.
The SC adds entries to the \ert when one of the nodes it is guarding are allocated as part of a new job; the SC deletes the entry when the job is over.

\myparagraph{Setting up and reserving non-TEE node} 
Given a manifest, CSP-MP asks for non-TEE node(s) from SC.
%
The SC first checks that the nodes are not part of any other job by scanning the \ert.
If the node(s) are unallocated, the SC starts the binding process.
Specifically, it sends attestation reports for itself and the non-TEE nodes allocated to the job. 
If successful, the user provisions a job-specific secret key $\mathcal{K}$ to the SC. 
The SC then adds $\mathcal{K}$ and the nodes associated with the job to the ERT.

\myparagraph{Communication: TEE-FDU and Non-TEE node}
The SC provides staging buffers where data can be encrypted and decrypted. 
For any data leaving the SC, the nodes move it to SC's buffer where it is encrypted with the destination's key before being sent out. 
Similarly, any data that is entering the SC (say, to one of the non-TEE nodes) arrives in the nodes memory mapped in the SC where it is decrypted with the destinations key. \Cref{fig:sc} shows the details of this mechanism. 

\myparagraph{Communication: Non-TEE under behind SC}
We do not employ authenticated encryption for data being exchanged between non-TEE nodes behind the same SC.
Instead we use the SC to perform access control using the ERT.
Nodes can communicate using the SC as in intermediary for performing data copies via an isolated staging buffer in the SC.
Alternatively, nodes can setup a zero-copy mechanism such that they can access each others memory directly. However, the SC has to mediate such a setup to ensure that only the nodes that belong to the same job can setup shared memory. Once setup, the SC need not be involved.

\myparagraph{Communication: Non-TEE nodes behind different SCs}
If a job requires nodes that are distributed across multiple SCs, the local SCs 
communicate over a secure channel using the job-specific key $\mathcal{K}$ to encrypt and decrypt the data. Since both SCs receive $\mathcal{K}$ during the setup and have valid \ert entries, this is feasible.
Note that SCs do not have to synchronize \ert entries or key material.

\myparagraph{Releasing node} 
When the tenant initiates a job termination, its primary CPU \fdu sends a termination command to the SC. %
Since the SC knows that the command came for a particular job, it resets the non-TEE nodes involved in the job. Specifically, its clears the internal state and removes the mapping from the \ert. 
Once reset, the SC notifies the MP that the node is available for future allocations.

\subsection{Securing TEE Nodes}
\label{sec:system:tee}

Unlike non-TEE nodes, TEE-nodes do not depend on an SC. 
TEE nodes provide \fdu{s} that are set up by the local SM and allows isolation of tenants by leveraging the on-device TEE primitives. 
Locally, the SM makes sure that the FDUs on the same node cannot access each other's memory. We use hardware root of trust and remote attestation of TEE nodes to ensure that such an isolation is trustworthy. 
Globally, we use authenticated encryption to protect data leaving the FDU. 
When two FDUs are explicitly communicating with each other, they can use a secure channel as shown in prior work~\cite{volos2018graviton}.
However, in our setting, we have to handle implicit communication such as DMA and MMIO. 

\myparagraph{Secure Implicit Communication}
On non-enclave platforms, the application communicates with a device (e.g., GPU) by writing to IO mapped regions, changes to which are reflected to the device. 
Alternatively, to send large amounts of data, the application can 
setup DMA transfers such that the device can read application memory directly, without involving the CPU.
In our solution, we have to protect the data when it is being transferred from the application to the device over an insecure network. 
We achieve this by ensuring that all data that leave's the application enclaves memory is protected with authenticated encryption. 
This requires monitoring all memory regions that are eventually accessed remotely.
One way to achieve this is to expect that the application will perform encryption decryption of data before it is exposed to remote access. 
This burdens developer effort and requires invasive application changes. 
Even with such changes, existing CPU TEEs do not allow non-enclave code (local or remote) to access enclave's private memory. 
Therefore, the enclave has to use untrusted address space for communication.
To address these challenges, we enable transparent authenticated encryption and copying (to non-enclave memory) of all enclave data that is being remotely accessed.

When applications use DMA or RDMA, they explicitly mark memory for DMA operations and initiate a data transfers. 
We can hook such events using a thin device driver inside the enclave. 
This way, we can delegate the monitoring of events and subsequent encryption and data copies to the driver, without changing the application. 

It is more challenging to support MMIO.
Applications maps the device memory into their own virtual address space and from then on simply use load and store operations, which are expected to reflect into device's memory.
Since there are no explicit MMIO events or syncs, we cannot hook such memory operations to perform encryption and copy the data outside the enclave. 
We observe that the device's MMIO-addressable memory range is fixed at boot. 
When the application sets up MMIO with the device, it has to map the devices MMIO address range into the applications address space. 
Since this setup is explicit, we can introduce a thin driver that hook the applications MMIO setup calls. 
Specifically, it marks the MMIO-mapped pages are inaccessible. 
This way, when the application performs MMIO accesses, it results in page faults. 
Our device driver can then register a handler for such page faults. During  execution, it can perform encryption-decryption and data copies between enclave and non-enclave memory region.

%% file: sections/case_studies_v3.tex
\section{Case-studies}
\label{sec:case_studies}

We identify four DSAs that are common across a number of public
clouds: GPU, FPGA, AI accelerator, and SSD. We use the existing TEE
proposals---Graviton~\cite{volos2018graviton} for GPU and ShEF~\cite
{zhao2021shef} for FPGA. We add TEE support to an AI accelerator and SSD.
Specifically, we identify three main hardware components to enable 
TEE support. 
First, an access control unit (\acu) enforces isolation across FDUs on
the same device by performing access checks on shared resources
(e.g., memory). On each resource access, the ACU knows the current
accessor and performs a look-up to ensure that it has the right
permissions.
Second, as a counterpart to the \acu, we maintain an \fdu{} mapping table
(\fmt) that keeps the mapping between each \fdu{} on the device to which the job is currently allocated.
The number of table entries is fixed depending on the number of \fdu{s} supported by the device. 
These table entries are populated and deleted by the device SM when an \fdu{} is assigned or removed from a job.
Finally, a memory protection engine (\mpe) that performs transparent
memory encryption and integrity protection for all the data entering
and leaving the device memory. 
We now explain the use of these components in peripheral TEEs.

\myparagraph{Existing Device TEEs} 
Graviton~\cite{volos2018graviton} instantiates a TEE for NVIDIA GPUs for isolated execution of kernels. 
It implements an \acu in a trusted security monitor on the GPU, which ensures isolation by
performing access checks on all commands submitted to the GPU from
the host as well as memory accesses from the GPU cores. 
Graviton tags all GPU pages with job IDs and performs filtering during runtime. Thus, it does not maintain a separate \fmt.
Each tenant uses a dedicated GPU kernel that acts as \mpe and uses an
authenticated encryption scheme. 
Graviton expects a hardware root of trust, such that the security monitor can generate attestation reports after a secure boot. 

ShEF does not allow multi-tenancy on the FPGA but instead provisions the entire FPGA to a single job. Thus, it does not need an \acu or a \fmt.
ShEF adds an \mpe called Shield to the user bitstreams which performs
encryption and decryption of all data leaving the tenant's
bitstream. 
SheF requires a hardware root-of-trust on the FPGA, such that a trusted security monitor can perform a secure boot and generate attestation reports. 
The tenant performs remote attestation and sets up a secure channel with the FPGA. 
During setup, ShEF guarantees the confidentiality and integrity of tenant's bitstreams. 
Summaries of Graviton and ShEF are in~\Cref{sec:appendix:existing_tees:fpga,sec:appendix:existing_tees:gpu}.

\begin{figure}[t]
  \centering
    \includegraphics[trim={0 16.5cm 22cm 0},clip,width=0.75\linewidth]{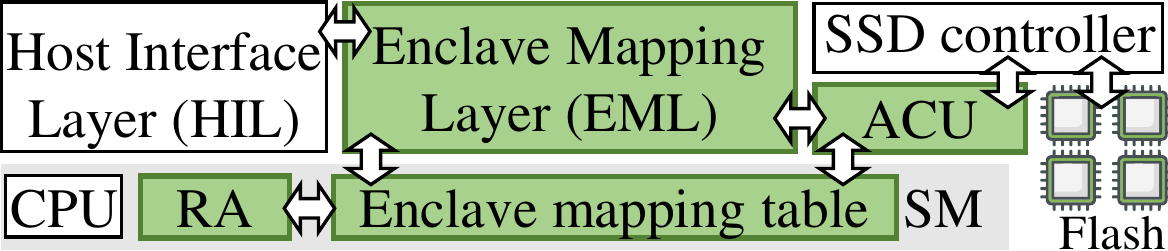}
    \caption{\textbf{SSD Case study}. A high-level model of an SSD. Additional hardware modules isolated enclave access are \green{highlighted}.
    }
    \spacesave
    \label{fig:eval:ssd_usecase}
\end{figure}
\subsection{Adding TEE Support for SSDs}
\label{sec:case_studies:ssd}

Tenants that share the same SSD can use tenant-specific keys to encrypt their data before it reaches the SSD, thus removing the need for a TEE. 
While straightforward, it is incompatible with modern data center deployments where storage computations are offloaded to the SSD accelerators, e.g., smart SSDs~\cite{smart_ssd}. 
These SSD accelerators require plain text access to the data. 
Additionally, recent research shows that a workloads such as data analytic~\cite{st1}, database query processing~\cite{st2,st3}, AI/ML\cite{st4}, map-reduce\cite{park2016storage}, and many commercial solutions~\cite{ngd,scaleflux} take advantage of in-storage computing. 
From these observations, we see that TEE primitives are necessary to support multi-tenancy on SSDs.

\myparagraph{A primer on SSD Architecture}
\Cref{fig:eval:ssd_usecase} shows a simplified version of the base model that contains the host interface layer (HIL) which implements an interface such as a PCIe subsystem to talk to the CPU, a SSD controller, and a CPU core (typically an ARM-M profile core) that runs the firmware of the SSD controller. The SSD controller bridges the HIL with the storage plane. For more details, refer to~\Cref{sec:appendix:ssd_arch}.

\myparagraph{Programming Model} 
SSDs expose four main block-level IO interfaces: \texttt{read}, \texttt{write}, \texttt{flush}, and \texttt{trim}, while more operations are supported at the file-system level. 
As our SSD-TEE works at the block level, the user can format a specific namespace with any file system, removing the need for application/OS changes. 
The existing device driver translates file operations(e.g., read, write, append) to block-level commands.

\myparagraph{TEE Support} 
We introduce hardware \acu and \mpe between the HIL and the SSD controller. 
This layer uses an \fmt programmed by the Secure Monitor, that runs on the on-board ARM core, to enforce isolation. 
On startup, the cloud provide divides the SSD into different \fdu{s}. 
The SM creates entries for each of the FDUs in the \fmt to register them. 
When an \fdu{} is assigned to a job, the SM looks up the \fmt, ensures that the \fdu{} is not allocated to any other job, and performs Remote Attestation using the SSD's hardware root-of-trust with the tenant. 
Using the RA process, the SM obtains a job-specific key that it programs into the \fmt. 
During runtime, the \acu intercepts all memory transactions sent from the HIL to the SSD controller and uses the \fmt to allow/deny transactions. 
Then, an \mpe uses the keys from the \fmt to encrypt/decrypt the data.

\begin{figure}[t]
  \centering
    \includegraphics[trim={0 13.9cm 21.5cm 0},clip,width=0.72\linewidth]{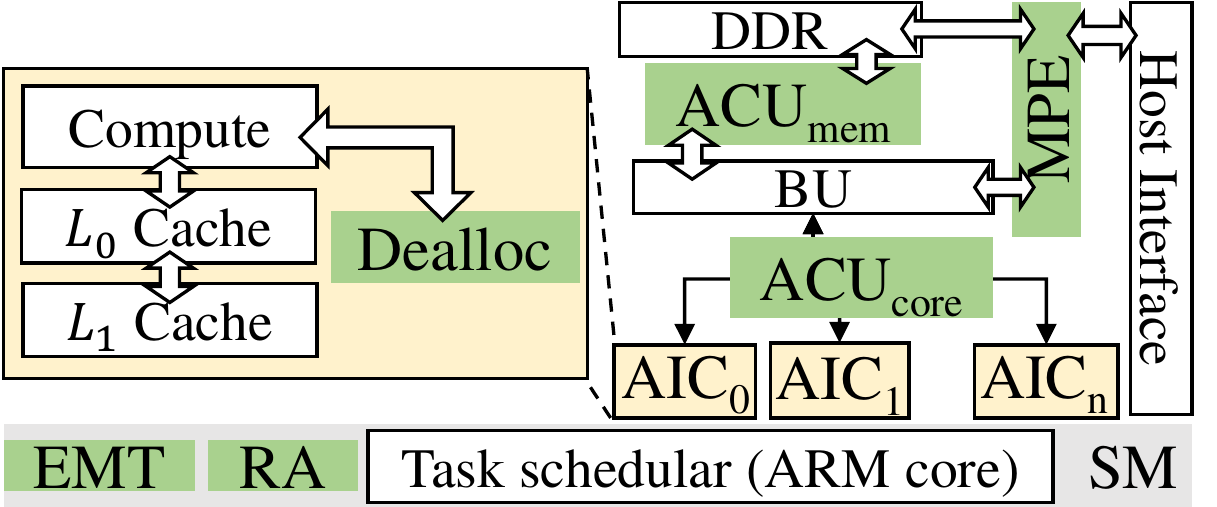}
    \caption{\textbf{AI Accelerator Case study}. An overview of the AI accelerator based on DaVinci~\cite{liao2019davinci,liao2021ascend} architecture. The additional hardware modules to enable TEE are \green{highlighted}.}
    \spacesave
    \label{fig:eval:ai_acc_usecase}
\end{figure}

\subsection{Adding TEE Support for AI Accelerators}
\label{sec:case_studies:ai_accelerator}

\myparagraph{Programming model} 
AI accelerator applications typically have a host counterpart that allocates accelerator memory for the computation. 
The host application copies data and commands to the AI accelerator using DMA before starting the execution. 
After completing execution, the accelerator writes back the results via DMA to the host memory. 

\myparagraph{A primer to an AI accelerator} 
\Cref{fig:eval:ai_acc_usecase} shows a simplified base model of AI
accelerator with $n$ cores based on the DaVinci architecture~\cite{liao2019davinci}. 
Each core is capable of running CNN layers and exposes a custom ISA. 
The cores are connected to DDR via the broadcast unit (BU). 
The BU collects memory requests from the cores and responses
from the memory subsystem and routes them to the correct endpoint.
Further, each core has a front end that acts as an interface for
communication with the other cores and BU. 
A detailed design is in~\cref{sec:appendix:acc}.

\myparagraph{TEE Support} 
At startup, the cloud provider divided the cores into multiple FDUs. 
When the FDU is
assigned to a job, the SM ensures that it is not already allocated by
looking up the \fmt. Then, it uses the hardware RoT to perform remote
attestation with the tenant to set up a job-specific shared key. The
SM programs this key into the \fmt. 
Because a job can span over multiple AI accelerator cores, it is
necessary to enable isolated inter-core communication. For this, we
use an \acuAICore that intercepts all inter-core traffic and uses
the \fmt to allow only the cores that belong to the same job to
communicate with each other. 
Next, to isolate accesses to the DDR memory we use an \acuAIMem to
filter all accesses based on the \fmt. Then, to encrypt all data
leaving the accelerator through the host interface we use an \mpe
that performs authenticated encryption using keys from the \fmt.
The secure deallocator (dealloc) tracks all tensors allocated to an FDU's memory space. 
When the SM wants to deallocate an FDU, the deallocator issues memory write commands to
zero out the \fdu reserved memory space on the DDR.

%% file: sections/security_analysis_v2.tex
\section{Security Analysis}
\label{sec:security_analysis}

\myparagraph{Attacks via untrusted Management Plane (CSP-MP)} 
The untrusted CSP-MP can violate the \job manifest (e.g., allocate wrong resources). 
However, the remote verifier can detect such discrepancies via the attestation report from \fdu{s} and SC (for non-TEE nodes). 
If CSP-MP allocates revoked nodes or nodes with old (possibly compromised) firmware, it will detected by remote attestation.
Nodes' local memory isolation ensures that the CSP-MP or hypervisor cannot access (read/write/clone) the enclave's private memory. 
The CSP-MP cannot compromise the secure channel between nodes and the remote verifier (both DMA and RDMA). 
The CSP-MP cannot allocate the same \fdu{} to two different tenants as the remote attestation locks a \fdu{} to a single tenant by the shared secret derived from the remote attestation.
The \ert on the SC prevents a non-TEE node from being allocated to two jobs simultaneously.

\myparagraph{Attacks through malicious configurations}  
Untrusted hypervisors or CSP can push malicious configurations to TEE or non-TEE nodes.
For TEE nodes, the remote verifier can verify the SM configurations and their integrity 
via the attestation report.
For non-TEE nodes, the SC reset them to their safe factory configuration before allocating them to a \fdu{}.
The attacker can always reset non-TEE nodes, but this will erase all the job-specific data and execution state without leaking secrets. 

\myparagraph{Attacks via untrusted \fdus/SM} 
The CSP can add compromised physical devices on the path (e.g., network devices without SMs) that can intercept traffic and trigger rogue DMAs and inject interrupts into nodes. 
This will not leak information or compromise the execution as the data is confidentiality and integrity-protected when it leaves a trusted \fdu. 
Remote attestation will detect the nodes running untrusted SMs (e.g., old vulnerable version, CSP modifies SM to generate insecure keys, SMs launch modified enclaves) and fail.

\myparagraph{Untrusted co-tenants} 
The node TEE's memory isolation prevents co-tenants from accessing a victim tenant's protected enclave private memory.
A tenant can attempt to perform MMIO or trigger DMA to devices that are part of the victim's job. 
Both these attacks are rendered harmless by encryption. 
Alternatively, the attacker can inject software-generated interrupts to the cores that are running the victim enclave. This could trick the enclave into starting a DMA with the attacker. 
This cannot compromise the victim because all data leaving the enclave's memory is encrypted and integrity protected. 

\myparagraph{Attacks via non-TEE nodes}
Malicious non-TEE nodes can try to access SC's buffers that store other node's data in plain text, say by reading the in/out buffers. The SC's access checks prevents such attacks. 
The nodes behind the same SC can try to communicate with each other directly (via MMIO or DMA), bypassing the SC. Since the nodes are only physically connected to the SC, such attacks would have to go through the SC's buses.  SC's bus-level access control checks prevent such unauthorized communication. 
We store all keys used to encrypt data leaving the non-TEE nodes in the SC's \ert. Malicious non-TEE nodes can try to read out the \ert, tamper with the \ert, or compromise the SC's execution. This is not possible as the SC runs in a privileged execution mode is inaccessible to other processes. Further, its data and memory is protected and is inaccessible to all other entities. 
%
The non-TEE nodes can try to send malicious MMIO and DMA requests to non-TEE nodes behind other SCs or other TEE FDUs. This does not compromise the confidentiality and integrity as all data is protected with authenticated encryption.

\myparagraph{Attacks via untrusted hypervisors}
The hypervisors are responsible for the node's memory management, loading the enclave's initial code, and setting up its memory regions. 
During remote attestation, we detect tampering of enclave code or additional initialization data and the \verifier will abort. 
If the hypervisor allocates enclave memory in the non-secure regions, the local SM will fail to launch the enclave. 
The local memory protection of the TEE ensures that the hypervisor cannot access the enclave's private memory region once it starts.
SM clears residual states from the memory if the hypervisor terminates an enclave, making sure that the previous data is not accessible to the hypervisor or other enclaves. 
The hypervisor can page out arbitrary enclave memory. However, TEE's memory isolation mechanisms ensure that the enclave pages are inaccessible to the hypervisor. 
Similarly, the TEE protect the enclave page table. This prevents the hypervisor from remapping the enclave's memory to non-protected memory. 

The hypervisor cannot inject DMAs to \fdus as the channel is confidentiality and integrity protected and will be detected by the trusted \fdu. 
Overlapping DMA regions does not compromise security as each pair of CPU \fdu{} and DSA \fdu{} (or SC) uses a unique job-specific key.
The hypervisor cannot reply to old attestation reports as the attestation challenge from the remote verifier ensures freshness.
The hypervisor can inject fake interrupts from devices to trigger a data transfer from the enclave's private memory to the device. However, encrypted DMA ensures no data is leaked.

\myparagraph{Physical Attacker}
Such an attacker cannot probe non-TEE nodes as they are confined in a tamper-resilient container; communication between non-TEE nodes connected to the SC never leaves the SC boundary.
The TEE node's memory is protected by local TEE features such as memory encryption and integrity protection.
A physical attacker cannot see or modify DMA authenticated and encrypted data between TEE nodes.
Communication between non-TEE nodes from different SCs is relayed through their local SCs that use authenticated encryption to protect the data.

%% file: sections/implementation_v3.tex
\section{Implementation}
\label{sec:implementation}

We implement the TEE support to AI accelerator and SSD devices in two ways:
RTL and high-level language. 
We implement the \acu, and \fmt in RTL as described in~\Cref{sec:case_studies} and then estimate the area, power, and timing costs.

\myparagraph{Functionality}
We augment an existing simulation of the device written in a high-level language (e.g., Python/C++),  with \acu, and \fmt.
For the \mpe, we implement the cryptographic operations in a high-level language , and measure \mpe{}'s execution time on the on-device ARM-v8 core.
For functional correctness, we ensure that the outputs from the original code with our modified code are same. 
Also, we use the functional model to collect execution traces that capture memory accesses. 

\myparagraph{Timing}
In each case, we use a device-specific cycle-accurate simulator timing estimation. 
In such a simulator, we integrate the timing behavior of our synthesized RTL logic into a known statistical model that computes the overall timing of the baseline RTL augmented with our custom RTL. 
We independently estimate the timing of individual operations from our RTL logic. We then update the simulator's timing model to consider these modified timings.
Given such a timing model and the execution traces, we estimate the overall execution time it would take 
to perform the same computation on a TEE-protected device.

For both devices, we implement a remote attestation unit using a HSM chip~\cite{tpm_chip, camenisch2017one} connected to the simulation platform over $I^2C$.

\subsection{Adding TEE Support: AI Accelerator}
\label{sec:implementation:ai}

We implement \acuAICore, \acuAIMem, \fmt, and enclave deallocator on each of the AI cores's front-end explained in~\Cref{sec:case_studies:ai_accelerator} as dedicated hardware units and the \mpe in SystemVerilog and Python. 
We add $200$ \loc in SystemVerilog for RTL for timing simulation and $650$ \loc in Python and shell script to existing $\sim$15K \loc for functional correctness. 
We use accelerator's onboard ARM core for the authenticated encryption of the \mpe. 

\myparagraph{Simulating the base model}
We use a \textit{cycle-accurate, event-based} 
simulator designed for Nvidia's cycle-accurate GPU that provides accurate timing and functional correctness~\cite{andri2022winograd}.
It models timing behavior, data movement (core, cache, and memory subsystem), and communication~\cite{villa2021need}. We use the timing model of an existing AI accelerator (Ascend's DaVinci AI core~\cite{liao2019davinci}).
To measure the timing impact of our RTL logic, we integrate it Nvidia's statistical model 
and event-based simulation.
We configure it for 800 MHz, and 150 AI core cycles latency to memory subsystem. These are standard settings typically used for accurate power and area estimations~\cite{766722}.
For verifying the correctness of calculations, we compare the execution output with 
real execution on a CPU.

We synthesize our SystemVerilog RTL implementation on a high-\textit{k} metal gate (HKMG) 28 nm CMOS technology, corresponding to a multi-VT standard cell library (supply voltage of 0.8 V, estimated in typical-typical (TT) corner~\cite{vlsi}). 
 
\myparagraph{DRAM} 
Computation on the AI accelerator involves execution on the core as well as interactions with DDR memory. The host copies data to accelerator's DRAM via DMA and the AI cores \texttt{load} and \texttt{store} data from the DRAM as well. 
Thus, to get complete timing estimates we use DRAMSim3~\cite{dramsim3,8999595}, a state-of-the-art cycle-accurate DRAM simulator.
We model the \texttt{DDR4-3200-x8} configuration and generate memory traces (example \cite{dramsim3_memory_trace})
of the DRAM subsystem.
We combine DRAM traces with the functional execution traces from the cores, and use the event-based simulation to get end-to-end timing.

\subsection{Adding TEE Support: SSD}
\label{sec:implemnetation:ssd}

\myparagraph{Simulating the base model} 
The base model of the SSD simulation is based on SimpleSSD~\cite{simplessd, 8031080}, a state-of-the-art cycle-accurate full-system simulator that closely simulates commercial SATA and NVMe SSDs with realistic performance characteristics. 

\myparagraph{Hardware changes} 
Our design does not necessitate changes to the SSD controller. 
We implement \acu, and \fmt in $259$ \loc of C++ directly to SimpleSSD's codebase ($\sim55K$). 
This allows us to perform functional simulation based on Gem5~\cite{gem5} that boots Linux and performs IO operations with the SSD. 
We implement these in $150$ \loc of SystemVerilog RTL synthesized and run place-and-route on 28nm and 150 \loc in python to verify the RTL using cocotb.
We use this implementation to estimate the timing costs per operation and we update the SimpleSSD's timing model accordingly. 
We also update the timing model with the costs for cryptographic operations performed by the \mpe. 
We use the enhanced timing model along with the functional simulation to get accurate execution-timing traces which we use to estimate the runtime overhead.

\subsection{Security Controller (SC)}
\label{sec:implementation:sc}

We implement a software SC prototype in C++. We use \texttt{libsodium} for AES256-GCM with Intel's AES-NI C intrinsic to accelerate AES, and Intel's AVX2 for fast memory copy (\texttt{\_mm256\_stream\_si256}). 
Our prototype implementation is single-threaded and handles a limited number of devices as the main bottleneck is AES-GCM bandwidth. 
It is straightforward to extend it to multiple cores of multi-channel memory to scale the performance with an increasing number of connected nodes.  
We add around $400$ LoC of wrapper code along with unmodified libsodium library and a runtime for communication between TEE and non-TEE nodes

%% file: sections/evaluation_v3.tex
\section{Evaluation}
\label{sec:eval}

\begin{table}[t]
\caption{Evaluation Platforms}
\spacesave
\begin{adjustbox}{width=\columnwidth,center}
\begin{tabular}{lll}\toprule
Evaluation & Hardware & OS/RT/Driver \\\midrule     
AI acc \& DRAM sim & 2$\times $Intel Xeon Gold 6240R, 1 TB DRAM & Ubuntu 20.04.5 LTS\\
AI acc & Huawei Atlas 300T (Ascend 910 SoC, 32 core) & Ascend runtime\\
SSD sim \& SC & Intel i9 11900K, 128 GB DRAM & Ubuntu 20.04.5 LTS\\
GPU & NVIDIA GeForce RTX 3080 & Nvidia driver \\
GPU \& FPGA host & Intel Xeon $2^{nd}$ Gen Scalable CPU & Ubuntu 20.04 LTS\\
FPGA ShEF & AWS F1 instances & AWS + Xilinx XDMA\\
FPGA local & Xilinx UltraScale+ VCU11 & Xilinx XDMA\\
\bottomrule
\end{tabular}
\end{adjustbox}
\spacesave
\label{tab:eval:test_bench}
 \end{table}

\myparagraph{Our Setup} 
\Cref{tab:eval:test_bench} describes the hardware platforms, OS, runtime, driver we use. 
From an experimental standpoint, the amount of time it took us to 
run AI accelerator simulation was high.
Specifically, an execution that 
requires $1$ second in real-clock time, takes us $\sim20$K seconds (~$\sim5.5$ hours) 
to simulate. 
Further, this rate increases linearly with the input data size. 
In comparison to the AI accelerator, the simulation for SSD is faster---$1$ second in 
real-clock time takes us $\sim24$ seconds to simulate. 
Booting up 
Linux takes $\sim25$ minutes on x86 and $\sim60$ minutes on ARM.
For evaluation, we select benchmarks and workloads that reflect the most popular real-world computations deployed on cloud accelerators.

\subsection{Case Study 1: AI Accelerator}
\label{sec:eval:acc}

\myparagraph{Synthetic stress test benchmarks} We develop two synthetic workloads: i) one that is \textit{memory-bound}, ii) and one that is \textit{compute-bound}. To build these workloads, we use  \texttt{Im2col}~\cite{im2col}, the dominant computation of the core responsible for CNN layers~\cite{wang2021optimization}. 
\Cref{tab:eval:case_study_1:synthetic} shows the number of cycles required for both memory and compute-bound applications.
We see that the memory-bound workload incurs significantly more overhead on the baseline than the compute-bound workload. 
We observe that the overhead incurred on memory-bound computation with protected access (P acc) is $\sim9\times$ higher than the corresponding compute-bound workload. 
This is expected as such memory bound computation engages the protection units (\acu and \mpe) more than compute-bound applications. 
The SDA unit's performance depends on the available memory bandwidth of a core and the memory subsystem.
The two computations use similarly sized tensors. Therefore, the latency of SDA is similar for both memory (1901 cycles) and compute-bound (2125 cycles) workloads.

\myparagraph{Real-world AI workloads experiment setup} 
 \Cref{tab:ai_accel} provides the results for loading the models into the accelerator, running the image inference on CIFAR-10($D_1$)~\cite{krizhevsky2009learning} ($60k$ $32\times32$ RGB 163MB), and ImageNet ILSVRC($D_2$)~\cite{russakovsky2015imagenet} (1.4M $224\times224$ RGB, 157GB) .

 \begin{table}[t]
     \caption{Effect of protected access ACU+MPE (P acc) and secure de-allocation (SDA) on AI Accelerator. 
     NP: non-protected access.}
     \spacesave
    \begin{adjustbox}{width=0.9\columnwidth,center}
     \begin{tabular}{lll?lll}\toprule
     \multicolumn{3}{c?}{Memory bound (Filter dim = 1)} & \multicolumn{3}{c}{Compute bound (Filter dim = 3)} \\
     Access type & Cycles & Overhead & Access type & Cycles & Overhead\\  \midrule
     
     NP acc & 5821 & Baseline & NP acc & 13552 & Baseline \\ 
     NP acc + SDA & 7722 & 32.65\% & NP acc + SDA & 15677 & 15.68\%\\
     P acc & 6688 & 14.89\% & P acc & 13782 & 1.7\%\\ 
     P acc + SDA & 8847 & 51.98\% & P acc + SDA & 16198 & 19.54\%\\
    \bottomrule
     \end{tabular}
     \end{adjustbox}
     \spacesave
     \label{tab:eval:case_study_1:synthetic}
 \end{table}
 
\myparagraph{Setup and Teardown}
We measure the time to transfer the AES256-GCM-protected compiled models and instantiate the \fmt with access control information. 
Decryption \& tag verification incur an average overhead of 12.9\% as compared to the baseline without these protections. 
The deallocator tracks all tensors during execution and only clears up tensor memory during teardown. 
Therefore, the teardown costs increase with the size of the model.
However, these costs are independent of the data set size because the input data are well as the results are written directly into the tensor memory 
E.g., SSD-VGG-16, a large model(380.2 MB), and Resnet34, a smaller model (105MB) have teardown costs of 8.08 ms and 2.23 ms respectively.

\begin{table*}[]
\centering
\caption{\textbf{AI Accelerator case study.} The table shows the effect of adding TEE support in the Ascend AI accelerator.\\
\scriptsize
\centering
\setlength{\tabcolsep}{2pt}
\begin{tabular}{l|l|l|l}
$L_{act}$: \#activation and model parameter &  $D_1$: DMA time of batch size $\times$ CIFAR-10 image(s) & Setup: DMA time of model (\%overhead) & Base: Unmodified accelerator\\
$L_{bin}$: \#model binary (\texttt{.om} file) & $D_2$:  DMA time of batch size $\times$ ImageNet image(s) & Teardown: SDA time after enclave exit & Secure: Base+protection units\\
\end{tabular}
}
\spacesave
\resizebox{\textwidth}{!}{%
\setlength{\tabcolsep}{4pt}
\input{tables/AI_table}
}
\spacesave
\label{tab:ai_accel}
\end{table*}

\myparagraph{Runtime data copy costs} To understand the impact of securing the DMA transfers between the host and the AI accelerator we measure the time taken to load state-of-the-art AI networks to the accelerator memory with and without AES256-GCM protection performed by the \mpe. 
\Cref{tab:ai_accel} shows the transfer time and overhead for the two datasets. 
We evaluate all models with the same 2 datasets. So, the average DMA transfer time for a given a data set and batch size is the same across models.

For a 4KB DMA transfer (after cache warm-up), AES256-GCM takes $\sim$1.47 $\mu$s on \textit{one} Intel i9 11900K CPU core at 5.2 GHz, and $\sim$3.52 $\mu$s on \textit{one} ARM v8 Cortex-A CPU core at 800 MHz. 
Therefore, we incur noticeable overheads for runtime data transfer that employs a large number of decryption operations. However, a single round of inference takes more time. 
E.g., For Resnet-34 and batch size = 16, ImageNet transfer takes $7.32$ ms and the inference takes $11.82$ ms. 
In these cases, we pipeline the data transfer and inference (initiating the next transfer while the previous is still in execution on the accelerator ) such that total execution time is only $11.82$ ms.

\begin{table}[t]
    \caption{\textbf{Multi-tenant scaling}. Micro-benchmark of load/store latency in cycles (\%slowdown) from AI cores to DDR. Sequential (S)/random (R) Read \& Write (4K) bandwidth (MB/s/tenant) (\%scaling) for Samsung Z-SSD.}
    \spacesave
    \begin{adjustbox}{width=\columnwidth,center}
    \setlength{\tabcolsep}{3pt}
\begin{tabular}{llll|llll}
\hline
          & \multicolumn{3}{c|}{AI accelerator}                                                                                    &                      & \multicolumn{3}{c}{SSD}                                                                                               \\
\#Tenants & \multicolumn{1}{c}{\cellcolor[HTML]{EFEFEF}1} & \multicolumn{1}{c}{2} & \multicolumn{1}{c|}{\cellcolor[HTML]{EFEFEF}4} & \multicolumn{1}{c}{} & \multicolumn{1}{c}{\cellcolor[HTML]{EFEFEF}1} & \multicolumn{1}{c}{2} & \multicolumn{1}{c}{\cellcolor[HTML]{EFEFEF}4} \\ \hline
Load      & \cellcolor[HTML]{EFEFEF}294                   & 307.1 (4.4)           & \cellcolor[HTML]{EFEFEF}316.5 (7.4)            & S/R Read             & \cellcolor[HTML]{EFEFEF}383/372               & 191(50)/186(50)       & \cellcolor[HTML]{EFEFEF}92(75)/93(75)         \\
Store     & \cellcolor[HTML]{EFEFEF}255.5                 & 257.7(0.7)            & \cellcolor[HTML]{EFEFEF}316.5(23.4)            & S/R Write            & \cellcolor[HTML]{EFEFEF}410/410               & 202(49)/205(50)       & \cellcolor[HTML]{EFEFEF}97(75)/97(75)         \\ \hline
\end{tabular}
    \end{adjustbox}
    \spacesave
    \label{tab:eval:tenant_scale}
\end{table}

\myparagraph{Results} 
To enable protected access (ACU+MPE) we see a worst-case overhead of $14.89$\% for the synthetic memory-bound workloads (P acc), but only 3.98\% for a real-world AI workload (Resnet-50, batch size = 1). 
This shows that real-world workloads do not exhibit worst-case performance as they are neither purely memory or compute bound. 
Additionally, real-world workloads maximize data reuse by caching, which reduces the number of \texttt{load} and \texttt{store} instructions issued by the cores. 
This is apparent in all the scenarios (CPU, Base accelerator and Secure i.e., Base+\acu) in~\Cref{tab:ai_accel} where the average running time for a single image decreases with increasing batch size. 
E.g., in Resnet34 runtime overhead decreases from 3.12\% (batch=1) to 0.05\% (batch=16). 

Resnet-34 and Resnet-50 have the same number of \texttt{load} and \texttt{store} operation despite different inference time as they have the exact same memory access pattern. 
However, Resnet-50 runs for more iterations, and in these iterations accesses cached data. This results in an increase in runtime but does not increase the number of \texttt{load} and \texttt{store} operations. 
For teardown, we observe a very small average overhead for $1.7\%$ when running inference on the Imagenet dataset. 
Additionally, we observe that running the inference on a CPU with $64$ logical cores is always slower  on average by a factor of $37.66\times$ than running it on the just $1$ core of the AI accelerator.

\myparagraph{Scalability}
We run real-world workloads with increasing number of tenants. As the tenants are pinned to specific AI cores until the enclaves exit, there are no context switch overheads.
Therefore, the bandwidth between the core and DDR could only be the bottleneck factor. 
This bandwidth can be saturated if all the tenants issue \texttt{load} and \texttt{store} simultaneously.
We see that the real-world workloads with $4$ tenants does not saturate the large memory bandwidth ($47$ GBps). 
Therefore, we do not observe any noticeable performance degradation for real-world AI workloads with $4$ number of simultaneous tenants.
To analyze the worst-case scenario, we run the synthetic memory-bound stress test for 1,2, and 4 concurrent tenants as shown in table ~\Cref{tab:eval:tenant_scale}. 
When the AI core does not perform any meaningful computation but continuously issues \texttt{load} and \texttt{store} operations, we see a slowdown of $\sim30\%$ with $4$ tenants.

\begin{figure}[t]
  \centering
   \includegraphics[trim={0 11.8cm 11cm 0},clip,width=0.85\linewidth]{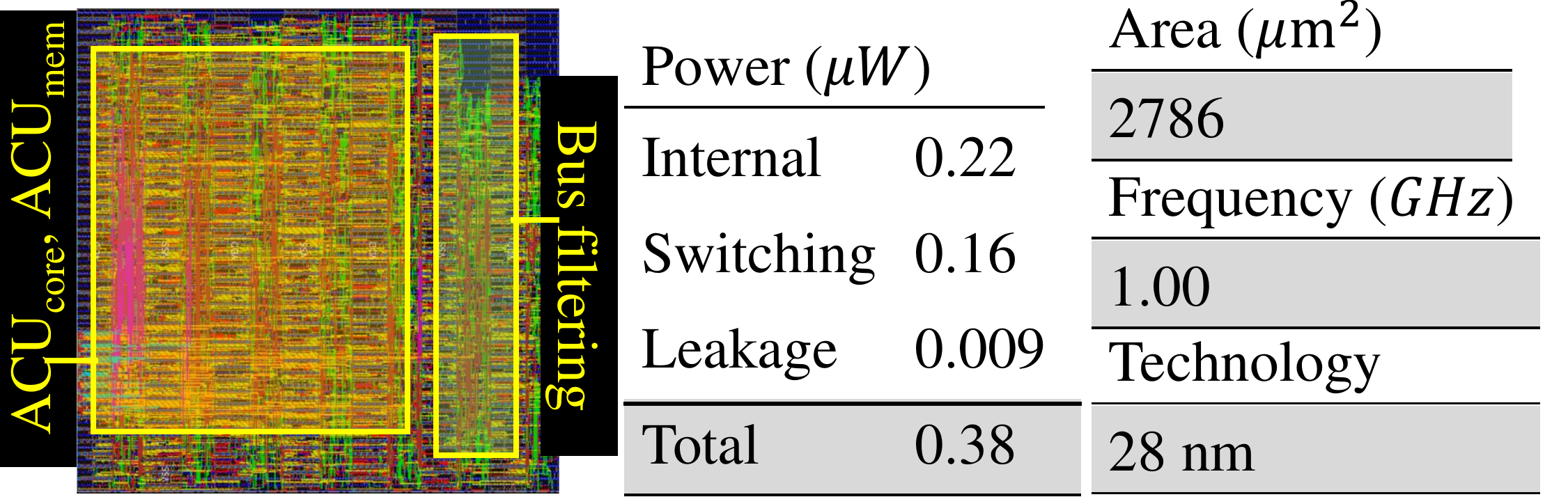}
    \caption{\textbf{Floor plan} of the synthesized hardware protection module on the AI accelerator alongside power~\cite{VLSI_power} and area breakdown.}
    \spacesave
    \label{fig:eval:floorplan}
\end{figure}

\myparagraph{Area overhead} \Cref{fig:eval:floorplan} shows the synthesized, placed \& routed floor plan of the hardware design that implements isolation mechanisms in the memory and the interface. 
The hardware only consumes 0.38 $\mu$W ($1.15\times 10^{-7}\%$) power and 1228 $mm^2$ ($0.2\%$) die area as compared to Ascend 910 which consumes 300W power and has a die area of 2786 $\mu m^2$.

\begin{table*}[h]
    \caption{\textbf{SSD case study.} Block IO (Read/Write) latency. ISO - \fdu isolation, ENC - AES256-GCM on six benchmarks: sequential reads (S-R), sequential writes (R-W), random reads (R-R), read-writes (R-W), sequential read-writes (S-RW), and random read-writes (R-RW)
    }
    \spacesave
    \begin{adjustbox}{width=2.1\columnwidth,center}
    \setlength{\tabcolsep}{4pt}

\input{tables/SSD_benchmark_V2}
    \end{adjustbox}
    \spacesave
    \label{tab:eval:case_study_2:overhead}
\end{table*}

\subsection{Case Study 2: SSD}
\label{sec:eval:ssd}
\myparagraph{SSD Models \& Workloads} We evaluate our design on $5$ types of SSD models provided in SimpleSSD.
We tested them with workloads from FIO, a standard tool for storage benchmark~\cite{fio} that perform six distinct disk operations (~\Cref{tab:eval:case_study_2:overhead}) on 2-16 GB of data. 
In FIO, we use the default 4KB block size (complies with Advanced Format standard~\cite{afk}). 
We perform a warmup for DRAM cache before our measurements, to reflect real-world workloads. 
We do not use any file system to avoid noises created by optimizations, however, we configure FIO to imitate LinuxFS behavior with an IO depth of 32.

\myparagraph{Performance Overheads} 
The performance of the SSD only depends on the block size, as it processes them one at a time in a streaming fashion. Thus, the total size of data transferred does not affect SSD's performance.
Therefore, we show the average latency overhead for 4K blocks in~\Cref{tab:eval:case_study_2:overhead}.
Given an SSD, we observe that all operations incur a fixed isolation and encryption overhead.
Further, as expected this latency is inversely proportional to the frequency of the ARM core on the SSD. 
A detailed specification of the SSDs used in our evaluation is shown in~\Cref{sec:appendix:ssd_arch}.
In the baseline, sequential read and sequential read-write are faster for than their random counterparts (e.g., on Intel 700, S-R takes 13.39 $\mu$s compared to $110.2\mu$s for R-R) as read operation take advantage of the caches. 
In contrast, write operations do not have cache advantages as the caches have a write-through policy to guarantee data persistence on the SSD. Therefore, the sequential and random writes have similar latency.  
The \mpe cannot use the caches as it performs AES-GCM before the data is stored in the cache. 
Hence, the overheads for AES-GCM as considerably high with the highest average of 31.93\% for S-RW.

\myparagraph{Scalability}
We instantiate 1, 2, and 4 simultaneous FIO~\cite{fio} benchmarks on the SSD and measure the read-write bandwidth. We observe that adding \acu and \mpe does not significantly reduce the bandwidth. E.g., the baseline bandwidth for sequential read for Samsung Z-SSD 800G SSD is $383.9$ MB/s and adding our protection does not change it as shown in~\cref{tab:eval:tenant_scale}.

\myparagraph{Area overhead} Compared to AI accelerator, the area overhead is significantly smaller due to simpler design. The hardware only take 650 $\mu m^2$ on the SSD controller.

\subsection{Case Study 3 \& 4: GPUs and FPGA}
\myparagraph{GPU MMIO and DMA} 
our design requires hooking on MMIO transfers (refer to~\Cref{sec:system:tee}) to protect them using AES-GCM. 
We use the Imagenet implementation and TinyImagenet dataset from the PyTorch library and perform training of different models as shown in~\Cref{tab:gpu-mmio}.
To understand when MMIO is used in different GPUs and to estimate the overhead for protecting them we use the \texttt{mmiotrace} utility in the Linux kernel to log all MMIO between application memory and GPU. 
We count the number of MMIO reads and writes to the driver-allocated application-specific memory-mapped region.  
We observe that MMIO is used to load model and data and there is no MMIO during the training and validation cycles. 
Specifically, $38\%$, $42\%$ and $20\%$ for setup, load data\& model, and teardown. 
Therefore, protecting these MMIO accesses will not incur any performance costs for the actual computation.
Further, we use the traces to estimate $1.89\%$ total overhead of adding AES256-GCM.

To capture DMA requests to and from the GPU, we use Nvidia's profiling utility \texttt{nsys}. 
We see that on the CPU, \texttt{cudaMemCopy} commands are issued before and after the kernels are executed.
To protect DMA we would require the driver to hook on these calls, encrypt the data, and copy it to non-secure memory. 
Similar to MMIO, this can be done before the kernels are launched. 
On a closer look, we see from the events captured on the GPU  that the data copies are performed in parallel with the computation. 
We can perform streaming encryption to maintain the speedups due to such streaming.

\begin{table}[]
\centering
\caption{\textbf{GPU case study:} The overhead\% on MIMO accesses shows the effect of adding TEE properties on the training process.}
\spacesave
\label{tab:gpu-mmio}
\resizebox{0.45\textwidth}{!}{%
\begin{tabular}{@{}lllllll@{}}
\toprule
Model                       & Batch & Setup  & LoadModel & LoadData & Teardown & Overhead\% \\ \midrule
\multirow{3}{*}{Resnet-32} & 32    & 106763 & 65736     & 57804    & 56398    & 2.28       \\
                            & 16    & 101683 & 64220     & 58435    & 57343    & 1.86       \\
                            & 8     & 109822 & 63689     & 57212    & 56272    & 1.54       \\ \midrule
\multirow{3}{*}{Resnet-50} & 32    & 114591 & 64970     & 57939    & 56523    & 2.12       \\
                            & 16    & 104517 & 63233     & 57503    & 56365    & 1.82       \\
                            & 8     & 108054 & 64596     & 56973    & 56011    & 1.26       \\ \midrule
Vgg16                       & 32    & 106536 & 61420     & 57182    & 55824    & 1.18       \\ \midrule
Squeezenet1\_1              & 32    & 111017 & 50923     & 56939    & 55623    & 2.54       \\ \midrule
Alexnet                     & 32    & 111971 & 64149     & 58329    & 57031    & 2.42       \\ \bottomrule
\end{tabular}%
}
\spacesave
\end{table}

\myparagraph{FPGA MMIO and DMA}
To analyze the costs of MMIO and DMA protection for FPGAs, we use DNNWeaver~\cite{dnnweaver}.
We execute on AWS F1~\cite{aws-f1} instances with and without the TEE feature provide by SheF~\cite{zhao2021shef}. 
We use \texttt{mmiotrace} and Xilinx DMA (\texttt{xdma}) driver to capture all MMIO and DMA accesses respectively. 
We observe the same data transfers sizes via DMA (1.41MB) and MMIO (16.91KB), with and without TEE features enabled. 
The computation performs small-sized MMIOs ($4$B) to initiate DMAs of size $4$KB to transfer models and data. 
During the FPGA execution, we observe no MMIO accesses. 
We see a average of $4344$ MMIOs and $362$ DMAs. 
To protect the transfers, we estimate similar overheads: $0.0033\%$ for MMIO and $0.0005$\% for DMA, with and without TEE features. 
We also observe that on average, the time between consecutive MMIO and DMA transfers are 10-20 $\mu$, 130 $\mu$s respectively. 
From our experiments, we know that the time to compute AES256-GCM with a block size of 4KB is only 1.47 $\mu$s as measured in~\Cref{sec:eval:acc}. Therefore, encryption will not become a bottleneck for the MMIO and DMA transfers. 

We implement SVD~\cite{svd} using Xilinx's Vitis libraries and run it on a local Xilinx vcu118 FPGA. 
The application uses \texttt{xdma} to transfer data to and from the FPGA. 
The FPGA computes the SVD of matrices ranging from $10\times10$ to $20000\times20000$, transfers data from $1.6$KB to $6.4$GB, causes $52$ and $9958$ MMIO,and $4$ to $766$ DMA operations respectively. 
Therefore, We observe that the number of MMIO and DMA are linearly proportional to the size of data. 
We use the MMIO and DMA traces to estimate the average latency introduced for protecting these accesses to be $0.04\%$ for MMIO and $0.002\%$ for DMA.

\subsection{Security Controller (SC)}
To evaluate the scaling limit of SC, we consider SC's individual operations. 
First, the SC performs AES256-GCM of data for the non-TEE nodes. 
The measured bandwidth of the native Intel AES-NI (using \texttt{libsodium}) AES256-GCM implementation is 2.59 GB/s/core. 
Next, we measure the bandwidth for copying data between encrypted and plaintext buffers (using Intel AVX2) to be 25 GB/s/core. 
Resnet-34, the fastest inference network that we evaluate, takes 1.2 ms/image/AI core. 
With the average size of 170KB/image in the ImageNet dataset, an AI core can be saturated with a image data rate of 138 MB/s from the CPU \fdu{}. 
Using this data, we estimate that the SC with a buffer size of 2.5 GB, on a single core can handle 19 CPU-\fdu{s} concurrently streaming image data. 
Adding jobs requires transferring the compiled trained model to the accelerator and is expensive, but it's a one-time cost per \job. 
With the largest model that we evaluate on (SSD300-VGG-16 which is 380 MB), SC on a single core can handle 6.97 jobs/sec without performance degradation. 

The number of nodes that can be connected to the SC and the buffer sizes required to support them, scales linearly with the number of CPU cores the SC uses. 
E.g., instead of a single core, we can deploy the SC on an Intel Xeon Gold processor with 48 cores as long as the memory system has enough bandwidth to support the SC buffers (171.5 GB/s for DDR4 8-channel 3200MHz).
Such a multi-core SC deployment theoretically can handle up to 912 nodes or 334 nodes/sec simultaneously for inference or starting up jobs respectively.

%% file: tables/AI_table.tex
\setlength{\extrarowheight}{0pt}
\addtolength{\extrarowheight}{\aboverulesep}
\addtolength{\extrarowheight}{\belowrulesep}
\setlength{\aboverulesep}{0pt}
\setlength{\belowrulesep}{0pt}
\begin{tabular}{lS[table-format=2]S[table-format=3]S[table-format=3]|S[table-format=3.1,round-mode=figures,round-precision=1]S[table-format=3.1,round-mode=figures,round-precision=4]|S[table-format=2.2,round-mode=places,round-precision=2]lS[table-format=1.2,round-precision=3]|S[table-format=2.2,round-precision=2]S[table-format=4]S[table-format=3.2,round-precision=4]S[table-format=4]|S[table-format=4.2,round-precision=2]S[table-format=2.2,round-mode=places,round-precision=2]S[table-format=2.2,round-mode=places,round-precision=2]l|S[table-format=7]S[table-format=6]}
\toprule
\multicolumn{6}{c|}{Model Information}                                                                                                                                                                                                                                                      & \multicolumn{3}{c|}{One time cost (ms)}                                                                      & \multicolumn{10}{c}{Runtime costs (On a single DaVinci core)}                                                                                                                                                                                                                                                                                                                                                                                               \\
\multicolumn{4}{c|}{Static Model Information}                                                                                                                                             & \multicolumn{2}{c|}{Compiled (MB)}                                                             & \multicolumn{2}{c}{Set up}                                                  & \multicolumn{1}{c|}{Teardown} & \multicolumn{2}{c}{Base DMA($\mu$s)}                                                      & \multicolumn{2}{c|}{Secure DMA ($\mu$s)}                                                   & \multicolumn{4}{c|}{Inference Runtime (ms)}                                                                                                                            & \multicolumn{2}{c}{\#Memory Access}                                                       \\
\hline
\multicolumn{1}{c}{Model} & \multicolumn{1}{c}{{\cellcolor[rgb]{0.945,0.945,0.945}}Batch} & \multicolumn{1}{c}{Res} & \multicolumn{1}{c|}{{\cellcolor[rgb]{0.937,0.937,0.937}}\# Layers} & \multicolumn{1}{c}{$L_{act}$} & \multicolumn{1}{c|}{{\cellcolor[rgb]{0.937,0.937,0.937}}$L_{bin}$} & \multicolumn{1}{c}{Base} & \multicolumn{1}{c}{{\cellcolor[rgb]{0.937,0.937,0.937}}Secure(\%)}   & \multicolumn{1}{c|}{Secure}   & \multicolumn{1}{c}{$D_1$} & \multicolumn{1}{c}{{\cellcolor[rgb]{0.937,0.937,0.937}}$D_2$} & \multicolumn{1}{c}{$D_1$} & \multicolumn{1}{c|}{{\cellcolor[rgb]{0.937,0.937,0.937}}$D_2$} & \multicolumn{1}{c}{CPU (64C)} & \multicolumn{1}{c}{{\cellcolor[rgb]{0.937,0.937,0.937}}Base} & \multicolumn{1}{c}{Secure} & Overhead                                   & \multicolumn{1}{c}{Load} & \multicolumn{1}{c}{{\cellcolor[rgb]{0.937,0.937,0.937}}Store}  \\
\hline
\multirow{3}{*}{Resnet-34} & {\cellcolor[rgb]{0.945,0.945,0.945}}1                         & 224                     & {\cellcolor[rgb]{0.937,0.937,0.937}}34                             & 2                           & {\cellcolor[rgb]{0.937,0.937,0.937}}103                          & 8.2                      & {\cellcolor[rgb]{0.937,0.937,0.937}}9.31 (13.5)  & 2.23                          & 4.47                      & {\cellcolor[rgb]{0.937,0.937,0.937}}106                       & 19                        & {\cellcolor[rgb]{0.937,0.937,0.937}}458                        & 15.29                         & {\cellcolor[rgb]{0.937,0.937,0.937}}1.206                    & 1.2437                     & {\cellcolor[rgb]{0.937,0.937,0.937}}3.12\% & 310382                   & {\cellcolor[rgb]{0.937,0.937,0.937}}4846                       \\
                           & {\cellcolor[rgb]{0.945,0.945,0.945}}8                         & 224                     & {\cellcolor[rgb]{0.937,0.937,0.937}}34                             & 2                           & {\cellcolor[rgb]{0.937,0.937,0.937}}103                          & 8.2                      & {\cellcolor[rgb]{0.937,0.937,0.937}}9.31 (13.5)  & 2.23                          & 35.76                     & {\cellcolor[rgb]{0.937,0.937,0.937}}848                       & 152                       & {\cellcolor[rgb]{0.937,0.937,0.937}}3664                       & 15.46                         & {\cellcolor[rgb]{0.937,0.937,0.937}}6.0726                   & 6.13                       & {\cellcolor[rgb]{0.937,0.937,0.937}}0.09\% & 2482146                  & {\cellcolor[rgb]{0.937,0.937,0.937}}39202                      \\
                           & {\cellcolor[rgb]{0.945,0.945,0.945}}16                        & 224                     & {\cellcolor[rgb]{0.937,0.937,0.937}}34                             & 2                           & {\cellcolor[rgb]{0.937,0.937,0.937}}103                          & 8.2                      & {\cellcolor[rgb]{0.937,0.937,0.937}}9.31 (13.5)  & 2.23                          & 71.52                     & {\cellcolor[rgb]{0.937,0.937,0.937}}1696                      & 304                       & {\cellcolor[rgb]{0.937,0.937,0.937}}7328                       & 15.52                         & {\cellcolor[rgb]{0.937,0.937,0.937}}11.755                   & 11.823                     & {\cellcolor[rgb]{0.937,0.937,0.937}}0.05\% & 4964290                  & {\cellcolor[rgb]{0.937,0.937,0.937}}78402                      \\
\hline
\multirow{3}{*}{Resnet-50} & {\cellcolor[rgb]{0.945,0.945,0.945}}1                         & 224                     & {\cellcolor[rgb]{0.937,0.937,0.937}}50                             & 3.5                         & {\cellcolor[rgb]{0.937,0.937,0.937}}108                          & 8.71                     & {\cellcolor[rgb]{0.937,0.937,0.937}}9.84 (12.9)  & 2.37                          & 4.47                      & {\cellcolor[rgb]{0.937,0.937,0.937}}106                       & 19                        & {\cellcolor[rgb]{0.937,0.937,0.937}}458                        & 22.11                         & {\cellcolor[rgb]{0.937,0.937,0.937}}1.836                    & 1.909                      & {\cellcolor[rgb]{0.937,0.937,0.937}}3.97\% & 310382                   & {\cellcolor[rgb]{0.937,0.937,0.937}}4846                       \\
                           & {\cellcolor[rgb]{0.945,0.945,0.945}}8                         & 224                     & {\cellcolor[rgb]{0.937,0.937,0.937}}50                             & 3.5                         & {\cellcolor[rgb]{0.937,0.937,0.937}}108                          & 8.71                     & {\cellcolor[rgb]{0.937,0.937,0.937}}9.84 (12.9)  & 2.37                          & 35.76                     & {\cellcolor[rgb]{0.937,0.937,0.937}}848                       & 152                       & {\cellcolor[rgb]{0.937,0.937,0.937}}3664                       & 23.13                         & {\cellcolor[rgb]{0.937,0.937,0.937}}10.706                   & 10.87                      & {\cellcolor[rgb]{0.937,0.937,0.937}}1.53\% & 2482146                  & {\cellcolor[rgb]{0.937,0.937,0.937}}39202                      \\
                           & {\cellcolor[rgb]{0.945,0.945,0.945}}16                        & 224                     & {\cellcolor[rgb]{0.937,0.937,0.937}}50                             & 3.5                         & {\cellcolor[rgb]{0.937,0.937,0.937}}108                          & 8.71                     & {\cellcolor[rgb]{0.937,0.937,0.937}}9.84 (12.9)  & 2.37                          & 71.52                     & {\cellcolor[rgb]{0.937,0.937,0.937}}1696                      & 304                       & {\cellcolor[rgb]{0.937,0.937,0.937}}7328                       & 23.29                         & {\cellcolor[rgb]{0.937,0.937,0.937}}24.195                   & 24.667                     & {\cellcolor[rgb]{0.937,0.937,0.937}}1.95\% & 4964290                  & {\cellcolor[rgb]{0.937,0.937,0.937}}78402                      \\
\hline
RetinaNet-RN50-fpn         & {\cellcolor[rgb]{0.945,0.945,0.945}}1                         & 800                     & {\cellcolor[rgb]{0.937,0.937,0.937}}50                             & 3.5                         & {\cellcolor[rgb]{0.937,0.937,0.937}}228.4                        & 18.11                    & {\cellcolor[rgb]{0.937,0.937,0.937}}20.38 (12.5) & 4.93                          & 4.47                      & {\cellcolor[rgb]{0.937,0.937,0.937}}106                       & 4.47                      & {\cellcolor[rgb]{0.937,0.937,0.937}}106                        & 329.56                        & {\cellcolor[rgb]{0.937,0.937,0.937}}30.515                   & 30.684                     & {\cellcolor[rgb]{0.937,0.937,0.937}}0.55\% & 4948128                  & {\cellcolor[rgb]{0.937,0.937,0.937}}278228                     \\
\hline
SSD300-VGG-16              & {\cellcolor[rgb]{0.945,0.945,0.945}}1                         & 300                     & {\cellcolor[rgb]{0.937,0.937,0.937}}32                             & 115.2                       & {\cellcolor[rgb]{0.937,0.937,0.937}}265                          & 29.7                     & {\cellcolor[rgb]{0.937,0.937,0.937}}33.47 (12.6) & 8.08                          & 4.47                      & {\cellcolor[rgb]{0.937,0.937,0.937}}106                       & 4.47                      & {\cellcolor[rgb]{0.937,0.937,0.937}}106                        & 37.39                         & {\cellcolor[rgb]{0.937,0.937,0.937}}6.539                    & 6.631                      & {\cellcolor[rgb]{0.937,0.937,0.937}}1.4\%  & 1199912                  & {\cellcolor[rgb]{0.937,0.937,0.937}}75176                      \\
\hline
UNet                       & {\cellcolor[rgb]{0.945,0.945,0.945}}1                         & 572                     & {\cellcolor[rgb]{0.937,0.937,0.937}}23                             & 8                           & {\cellcolor[rgb]{0.937,0.937,0.937}}119                          & 9.92                     & {\cellcolor[rgb]{0.937,0.937,0.937}}11.24 (13.3) & 2.7                           & 4.47                      & {\cellcolor[rgb]{0.937,0.937,0.937}}106                       & 4.47                      & {\cellcolor[rgb]{0.937,0.937,0.937}}106                        & 224.89                        & {\cellcolor[rgb]{0.937,0.937,0.937}}23.379                   & 23.479                     & {\cellcolor[rgb]{0.937,0.937,0.937}}0.42\% & 3077286                  & {\cellcolor[rgb]{0.937,0.937,0.937}}187302                     \\
\hline
\multirow{3}{*}{YOLO V3}   & {\cellcolor[rgb]{0.945,0.945,0.945}}1                         & 256                     & {\cellcolor[rgb]{0.937,0.937,0.937}}106                            & 50                          & {\cellcolor[rgb]{0.937,0.937,0.937}}237                          & 22.42                    & {\cellcolor[rgb]{0.937,0.937,0.937}}25.30(12.86) & 6.1                           & 4.47                      & {\cellcolor[rgb]{0.937,0.937,0.937}}106                       & 19                        & {\cellcolor[rgb]{0.937,0.937,0.937}}458                        & 92.9                          & {\cellcolor[rgb]{0.937,0.937,0.937}}3.606                    & 3.747                      & {\cellcolor[rgb]{0.937,0.937,0.937}}3.91\% & 150145                   & {\cellcolor[rgb]{0.937,0.937,0.937}}9857                       \\
                           & {\cellcolor[rgb]{0.945,0.945,0.945}}1                         & 416                     & {\cellcolor[rgb]{0.937,0.937,0.937}}106                            & 50                          & {\cellcolor[rgb]{0.937,0.937,0.937}}237                          & 22.42                    & {\cellcolor[rgb]{0.937,0.937,0.937}}25.30(12.86) & 6.1                           & 4.47                      & {\cellcolor[rgb]{0.937,0.937,0.937}}106                       & 19                        & {\cellcolor[rgb]{0.937,0.937,0.937}}458                        & 1386.19                       & {\cellcolor[rgb]{0.937,0.937,0.937}}7.591                    & 7.709                      & {\cellcolor[rgb]{0.937,0.937,0.937}}1.55\% & 396475                   & {\cellcolor[rgb]{0.937,0.937,0.937}}26027                      \\
                           & {\cellcolor[rgb]{0.945,0.945,0.945}}8                         & 256                     & {\cellcolor[rgb]{0.937,0.937,0.937}}106                            & 50                          & {\cellcolor[rgb]{0.937,0.937,0.937}}237                          & 22.42                    & {\cellcolor[rgb]{0.937,0.937,0.937}}25.30(12.86) & 6.1                           & 35.76                     & {\cellcolor[rgb]{0.937,0.937,0.937}}848                       & 152                       & {\cellcolor[rgb]{0.937,0.937,0.937}}3664                       & 3400.99                       & {\cellcolor[rgb]{0.937,0.937,0.937}}19.23                    & 19.45                      & {\cellcolor[rgb]{0.937,0.937,0.937}}1.44\% & 1201153                  & {\cellcolor[rgb]{0.937,0.937,0.937}}78849                      \\
\bottomrule
\end{tabular}

%% file: tables/SSD_benchmark_V2.tex
\begin{tabular}{cllS[table-format=3.2]S[table-format=1.4]S[table-format=3.2]S[table-format=4.2]S[table-format=1.4]S[table-format=3.2]S[table-format=3.2]S[table-format=1.3]S[table-format=2.2]S[table-format=4.2]S[table-format=1.4]S[table-format=3.2]S[table-format=4.2]S[table-format=1.3]S[table-format=2.2]S[table-format=4.2]S[table-format=1.4]S[table-format=2.3]} 
\toprule
                                &                         &                                                                                          & \multicolumn{3}{c}{{\cellcolor[rgb]{0.827,0.827,0.827}}S-R}                                                                                                                                                                                                                                                                                                                                                                                                                        & \multicolumn{3}{c}{S-W}                                                                                                                                                                                                                 & \multicolumn{3}{c}{{\cellcolor[rgb]{0.827,0.827,0.827}}R-R}                                                                                                                                                                                                                                                                                                                                                                                                        & \multicolumn{3}{c}{R-W}                                                                                                                                                                                                                 & \multicolumn{3}{c}{{\cellcolor[rgb]{0.827,0.827,0.827}}S-RW}                                                                                                                                                                                                                                                                                                                                                                                                       & \multicolumn{3}{c}{R-RW}                                                                                                                                                                                                                 \\ 
\hline
                                & \multicolumn{1}{c}{SSD} & \multicolumn{1}{c}{OP cost}                                                              & \multicolumn{1}{c}{{\cellcolor[rgb]{0.827,0.827,0.827}}\begin{tabular}[c]{@{}>{\cellcolor[rgb]{0.827,0.827,0.827}}c@{}}Base\\$\mu$s\end{tabular}} & \multicolumn{1}{c}{{\cellcolor[rgb]{0.827,0.827,0.827}}\begin{tabular}[c]{@{}>{\cellcolor[rgb]{0.827,0.827,0.827}}c@{}}ISO\\\%\end{tabular}} & \multicolumn{1}{c}{{\cellcolor[rgb]{0.827,0.827,0.827}}\begin{tabular}[c]{@{}>{\cellcolor[rgb]{0.827,0.827,0.827}}c@{}}ENC\\\%\end{tabular}} & \multicolumn{1}{c}{\begin{tabular}[c]{@{}c@{}}Base\\$\mu$s\end{tabular}} & \multicolumn{1}{c}{\begin{tabular}[c]{@{}c@{}}ISO\\\%\end{tabular}} & \multicolumn{1}{c}{\begin{tabular}[c]{@{}c@{}}ENC\\\%\end{tabular}} & \multicolumn{1}{c}{{\cellcolor[rgb]{0.827,0.827,0.827}}\begin{tabular}[c]{@{}>{\cellcolor[rgb]{0.827,0.827,0.827}}c@{}}Base\\$\mu$s\end{tabular}} & \multicolumn{1}{c}{{\cellcolor[rgb]{0.827,0.827,0.827}}\begin{tabular}[c]{@{}>{\cellcolor[rgb]{0.827,0.827,0.827}}c@{}}ISO\\\%\end{tabular}} & \multicolumn{1}{c}{{\cellcolor[rgb]{0.827,0.827,0.827}}\begin{tabular}[c]{@{}>{\cellcolor[rgb]{0.827,0.827,0.827}}c@{}}ENC\\\%\end{tabular}} & \multicolumn{1}{c}{\begin{tabular}[c]{@{}c@{}}Base\\$\mu$s\end{tabular}} & \multicolumn{1}{c}{\begin{tabular}[c]{@{}c@{}}ISO\\\%\end{tabular}} & \multicolumn{1}{c}{\begin{tabular}[c]{@{}c@{}}ENC\\\%\end{tabular}} & \multicolumn{1}{c}{{\cellcolor[rgb]{0.827,0.827,0.827}}\begin{tabular}[c]{@{}>{\cellcolor[rgb]{0.827,0.827,0.827}}c@{}}Base\\$\mu$s\end{tabular}} & \multicolumn{1}{c}{{\cellcolor[rgb]{0.827,0.827,0.827}}\begin{tabular}[c]{@{}>{\cellcolor[rgb]{0.827,0.827,0.827}}c@{}}ISO\\\%\end{tabular}} & \multicolumn{1}{c}{{\cellcolor[rgb]{0.827,0.827,0.827}}\begin{tabular}[c]{@{}>{\cellcolor[rgb]{0.827,0.827,0.827}}c@{}}ENC\\\%\end{tabular}} & \multicolumn{1}{c}{\begin{tabular}[c]{@{}c@{}}Base\\$\mu$s\end{tabular}} & \multicolumn{1}{c}{\begin{tabular}[c]{@{}c@{}}ISO\\\%\end{tabular}} & \multicolumn{1}{c}{\begin{tabular}[c]{@{}c@{}}ENC\\\%\end{tabular}}  \\ 
\hline
\multicolumn{1}{l}{}            & Intel 700 400 G         & \begin{tabular}[c]{@{}l@{}}ISO=5ns\\ENC=10.24 $\mu$s\end{tabular}     & {\cellcolor[rgb]{0.827,0.827,0.827}}13.39                                                                                                                            & {\cellcolor[rgb]{0.827,0.827,0.827}}0.074                                                                                                                    & {\cellcolor[rgb]{0.827,0.827,0.827}}76.42                                                                                                    & 9.35                                                                                        & 0.1                                                                 & 109.4                                                               & {\cellcolor[rgb]{0.827,0.827,0.827}}110.2                                                                                                                            & {\cellcolor[rgb]{0.827,0.827,0.827}}0.009                                                                                                    & {\cellcolor[rgb]{0.827,0.827,0.827}}9.29                                                                                                     & 9.35                                                                                        & 0.1                                                                 & 109.4                                                               & {\cellcolor[rgb]{0.827,0.827,0.827}}58.26                                                                                                                            & {\cellcolor[rgb]{0.827,0.827,0.827}}0.017                                                                                                    & {\cellcolor[rgb]{0.827,0.827,0.827}}17.57                                                                                                    & 120.36                                                                                      & 0.008                                                               & 8.5                                                                  \\
\multirow{2}{*}{\rotatebox[origin=c]{90}{NVMe}} & Samsung Z-SSD 400G      & \begin{tabular}[c]{@{}l@{}}ISO=5ns \\ENC=10.24 $\mu$s\end{tabular}    & {\cellcolor[rgb]{0.827,0.827,0.827}}4.4                                                                                                                              & {\cellcolor[rgb]{0.827,0.827,0.827}}0.113                                                                                                                    & {\cellcolor[rgb]{0.827,0.827,0.827}}115.89                                                                                                   & 6.22                                                                                        & 0.080                                                               & 82.3                                                                & {\cellcolor[rgb]{0.827,0.827,0.827}}13.28                                                                                                                            & {\cellcolor[rgb]{0.827,0.827,0.827}}0.037                                                                                                    & {\cellcolor[rgb]{0.827,0.827,0.827}}38.54                                                                                                    & 6.22                                                                                        & 0.080                                                               & 82.3                                                                & {\cellcolor[rgb]{0.827,0.827,0.827}}10.14                                                                                                                            & {\cellcolor[rgb]{0.827,0.827,0.827}}0.049                                                                                                    & {\cellcolor[rgb]{0.827,0.827,0.827}}50.45                                                                                                    & 11.55                                                                                       & 0.043                                                               & 44.32                                                                \\
                                & Samsung 983DCT 1.92T    & \begin{tabular}[c]{@{}l@{}}ISO=5ns \\ENC=10.24 $\mu$s\end{tabular}    & {\cellcolor[rgb]{0.827,0.827,0.827}}12.39                                                                                                                            & {\cellcolor[rgb]{0.827,0.827,0.827}}0.04                                                                                                                     & {\cellcolor[rgb]{0.827,0.827,0.827}}41.29                                                                                                    & 8.7                                                                                         & 0.056                                                               & 58.27                                                               & {\cellcolor[rgb]{0.827,0.827,0.827}}27.74                                                                                                                            & {\cellcolor[rgb]{0.827,0.827,0.827}}0.018                                                                                                    & {\cellcolor[rgb]{0.827,0.827,0.827}}18.45                                                                                                    & 8.78                                                                                        & 0.056                                                               & 58.27                                                               & {\cellcolor[rgb]{0.827,0.827,0.827}}17.78                                                                                                                            & {\cellcolor[rgb]{0.827,0.827,0.827}}0.028                                                                                                    & {\cellcolor[rgb]{0.827,0.827,0.827}}28.79                                                                                                    & 83.17                                                                                       & 0.006                                                               & 6.155                                                                \\ 
\hline
\multirow{2}{*}{\rotatebox[origin=c]{90}{SATA}} & Samsung 850Pro 256G     & \begin{tabular}[c]{@{}l@{}}ISO=10ns \\ENC=10.24 $\mu$s\end{tabular}   & {\cellcolor[rgb]{0.827,0.827,0.827}}83.17                                                                                                                            & {\cellcolor[rgb]{0.827,0.827,0.827}}0.006                                                                                                                    & {\cellcolor[rgb]{0.827,0.827,0.827}}6.15                                                                                                     & 415.9                                                                                       & 0.002                                                               & 2.46                                                                & {\cellcolor[rgb]{0.827,0.827,0.827}}412.87                                                                                                                           & {\cellcolor[rgb]{0.827,0.827,0.827}}0.002                                                                                                    & {\cellcolor[rgb]{0.827,0.827,0.827}}2.48                                                                                                     & 419.72                                                                                      & 0.009                                                               & 9.29                                                                & {\cellcolor[rgb]{0.827,0.827,0.827}}419.56                                                                                                                           & {\cellcolor[rgb]{0.827,0.827,0.827}}0.023                                                                                                    & {\cellcolor[rgb]{0.827,0.827,0.827}}2.44                                                                                                     & 416.75                                                                                      & 0.0023                                                              & 2.45                                                                 \\
                                & Intel 535 240G          & \begin{tabular}[c]{@{}l@{}}ISO=13.33ns\\ENC=10.24 $\mu$s\end{tabular} & {\cellcolor[rgb]{0.827,0.827,0.827}}560.1                                                                                                                            & {\cellcolor[rgb]{0.827,0.827,0.827}}0.0023                                                                                                                   & {\cellcolor[rgb]{0.827,0.827,0.827}}2.43                                                                                                     & 1875                                                                                        & 0.0007                                                              & 0.72                                                                & {\cellcolor[rgb]{0.827,0.827,0.827}}727.4                                                                                                                            & {\cellcolor[rgb]{0.827,0.827,0.827}}0.002                                                                                                    & {\cellcolor[rgb]{0.827,0.827,0.827}}2.43                                                                                                     & 2063                                                                                        & 0.0006                                                              & 0.66                                                                & {\cellcolor[rgb]{0.827,0.827,0.827}}~1017                                                                                                                            & {\cellcolor[rgb]{0.827,0.827,0.827}}0.001                                                                                                    & {\cellcolor[rgb]{0.827,0.827,0.827}}1.34                                                                                                     & 1184                                                                                        & 0.001                                                               & 1.15                                                                 \\
\bottomrule
\end{tabular}

%% file: sections/related_work_v2.tex
\section{Related Work}
\label{sec:relatedWorks}

Building a scalable infrastructure for multi-tenant cloud has been a challenging distributed systems problem. 
Prior works have shown solutions~\cite{shen2011cloudscale,yu2015network} that are now deployed in production. 
However, they assume a trusted cloud service provide and only protected against a software attacker executing as a malicious tenant.
Our work focuses on stronger adversary model: untrusted cloud provider and a physical attacker.

\myparagraph{Device TEEs} 
Existing proposals enable TEE on devices such as
GPUs~\cite{volos2018graviton, hunt2020telekine, jang2019HIX, ng2021goten, kwon2019zerokernel}, FPGAs~\cite{zhao2021shef,zeitouni2021trusted,oh2021meetgo}, SSDs~\cite{kang2021iceclave}, and other accelerators~\cite{vaswani2022confidential}.
They focus solely on adding TEE-features so as to create hardware-enabled enclaves on the accelerator---either for multi-tenancy or for single-tenancy. 
They assume that the device is connected to a single physical host
machine that may optionally have TEE support. 
They primarily achieve isolation from the attacker-controlled driver,
OS, and hypervisor.
Some device TEE solutions additionally remove the device driver from
the TCB and provide protection against a physical adversary. 
The only known commercially announced non-CPU TEE is NVIDIA's upcoming
H100~\cite{h100} which enables secure isolated instances (known as
MIG~\cite{mig}) in the GPU using proprietary GPU firmware. 
Additionally, prior works use cryptographic primitives like homomorphic encryption~\cite{lopez2012fly}, build privacy-preserving machine-learning frameworks~\cite{tan2021cryptgpu,hesamifard2018privacy}, and oblivious multi-party computation for data-centers~\cite{ohrimenko2016oblivious}.

\myparagraph{Heterogeneous TEE} 
Prior proposals enable TEEs across a CPU-TEE and a device-TEE physically connected to each other. 
SGX-FPGA~\cite{xia2021sgx} proposes a custom TEE for FPGA. 
It uses a physical unclonable function (PUF) on the FPGA for hardware root of
trust. The FPGA is managed by a trusted SM that
communicates with the SGX enclave over an encrypted channel. 
HIX~\cite{jang2019HIX} achieves similar protection for GPUs where
an SGX enclave communicates with the GPU enclave over encrypted
shared memory using HIX trusted driver. 
Hector-V~\cite{nasahl2021hector} provides a switch wrapper configured
during boot time by the CPU-SM. It acts as a filter for peripherals
on the AXI4-lite bus to drop request coming from untrusted code.
Another line of research has shown the feasibility of doing enclaved
execution on unmodified devices by extending the CPU-TEE protection.
They leverage MMU protection to protect memory-mapped devices and
perform bus-level isolation since devices are physically connected to
a TEE host~\cite{schneider2022composite, bahmani2020cure, cronus}.

Our solution is compatible with these device-TEEs, but does not
require the devices to be directly connected to CPU-TEE hosts. 
We use the insights from above proposals to build the
TEE-support for AI accelerator and SSD, by adding protection units
(\acu, \fmt, \mpe) and relying on security monitors. 
Our SC-based access control for distributed setting over a network
mimics the bus-level filtering for a centralized setting where CPU
and devices are connected a bus.

\myparagraph{Distributed TEE} 
HETEE~\cite{zhu2020hetee} addresses a similar problem as ours, but in a different setting. 
It assumes non-TEE devices that makes it more general. 
However, it also assumes that devices are connected to a centralized security controller (SC) over a PCIe switch in a physical attack-proof container. 
This limits the scalability of HETEE to a single rack and $\sim 60$ nodes. 
In contrast, our design carefully avoids a centralized SC design ($S_0$ in~\Cref{fig:design_decisions} in~\Cref{sec:design_space:potential_designs}). 
Nonetheless, at least at the rack-level, our solution may look similar to HETEE.
HETEE's centralized SC solution leads to bottlenecks even within the rack. 
First, it does not allow multi-tenancy on any nodes, since they do not have TEE support. We allow multi-tenancy on TEE-nodes. 
Second, the SC has to perform encryption decryption operations for all data between 60 nodes. 
Hence, they employ trusted proxy nodes to evenly distribute the cryptographic operations. 
We avoid this problem by protecting the non-TEE nodes by the SC's access-control checks, 
encryption decryption is only required when communicating with TEE nodes. 

\myparagraph{Commercial confidential cloud solutions} 
Several CSPs offer accelerators as a unit~\cite{aws-f1, azure-fpga} or as a packaged service~\cite{huawei-ai, google-ai}. 
However, none of them are TEE-protected as of yet. 
CPU and hypervisor-based TEEs have become prominent. E.g., SGX \& SEV-based Microsoft Azure confidential
cloud~\cite{azure_confidential_cloud}, hypervisor-based Amazon
Nitro~\cite{AWS_nitro_enclave}, SEV-based Google confidential
GTK~\cite{google_confidential_computing}, and SGX-based IBM
confidential cloud~\cite{ibm_confidential_cloud}.
Azure announced confidential computing on NVIDIA A100s GPUs that allows isolated computation on a dedicated GPU connected to a AMD SEV protected CPU-TEE.

%% file: sections/conclusion.tex
\section{Conclusion}
\label{sec:conclusion}

We present a data-center design that provides enclaved execution
guarantees over a mix of TEE and non-TEE devices. Our
hybrid design pays off by scaling across several nodes as well as
being future-ready for TEE-enabled devices. 
Our evaluation shows that the performance costs of such a design are modest and can be
reduced further with fine-tuning. 
We envision that our insight will serve as guiding principles for cloud-scale
solutions for confidential computing beyond CPU execution.


%% file: sections/appendix.tex
\section{Appendix}
\label{sec:appendix}

\subsection{Enclave Manifest}
\label{sec:appendix:manifest}

\begin{figure}[h]
\begin{lstlisting}[language=json,firstnumber=1]
{"Enclave" : "Enclave1",
"Enclave Vendor" : "Vendor1",
"Version" : "X.YZ",
"Resource" : {
  {"Resource type : CPU", "Policies" : {"no-HT"}
  "Cores" : 2, "Memory" : 256M },
  {"Resource type : AI_Accelerator",
  "Policies : {"memIsolation", "cachePartitioned"},
  "Cores" : 20, "Memory" : 16G}
 "SHA-3":"0xe3bae2603b...", "Sig":"18fd2fa796..."}
\end{lstlisting}
    \caption{Example manifest file where the 
    the programmer specifies the number, size, and type of each \fdu as well as the code to be executed on each \fdu.
    }
    \label{fig:manifest}
\end{figure}

To launch an enclave with trusted \fdu{s} and non-TEE nodes, the programmer uses a CSP provided interface, to specify the number, size, and type of each \fdu and non-TEE node in a manifest file. Along with this, the programmer also provides the code to be executed on each \fdu{} and non-TEE node. One example manifest is provide in~\Cref{fig:manifest}. The manifest is signed by the private key of the enclave developer to ensure that the integrity of the enclave manifest is preserved. The remote verifier can use the manifest to verify the combined attestation report (example in~\Cref{fig:acc_report}) to ensure that the allocations of the \fdu{s} and non-TEE node are correct.

\subsection{DSA Configuration Attestation}
\label{sec:appendix:attestation}

One potential way to do the DSA attestation is to use software-based attestation. However, a number of drawbacks in software-attestation (refer to~\ref{sec:appendix:sw-attestation}) makes it not suitable in our case.
The remote verifier starts the remote attestation process by sending unique attestation challenges to each of the local SMs. This process can be automated by a user-side software module that can generate unique challenges. The attestation challenges ensure that the attacker cannot reuse old attestation reports.
First, the CPU SM generates the attestation report of the enclave and public-private key pair for the enclave. 
At the same time, the CPU SM forwards the DSA-specific challenges to the allocated \fdu{s}. Alongside traditional remote attestation, the DSAs leverage \emph{property-based attestation.} (PBA)~\cite{sadeghi2004property,chen2006protocol} that ensures if a DSA has certain security properties.
We opt for PBA as the DSAs are highly programmable and attacker-controlled CSP or hypervisors can push malicious configurations to the DSA to compromise user enclave data.
Security properties such as cache partitioning, disabled debug port etc. 
can enforced by the DSA-SM. Therefore, the untrusted software stack cannot manipulate the device configurations. During the measured boot process, the local SM updates the platform configuration registers (PCR) with a pre-determined configuration value that reflects the correct execution of a property as the following:

\begin{figure}[t]
\begin{lstlisting}[language=ccode,firstnumber=1]
if(debug_port == false) //check for debug port status
    TPM2_pcrExtend(23, 0xb2..);//extend 23rd PCR bank
else TPM2_pcrExtend(23, 0xf5..);
\end{lstlisting}
\caption{Committing predefined hash value to secure non-volatile memory on the hardware security module (HSM) on DSA when certain security properties are enforced in the firmware during startup.}
\label{fig:pcr_value}
\end{figure}

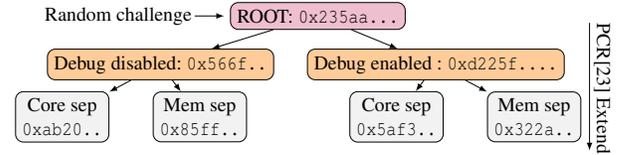
\begin{figure}[h]
\centering
\begin{adjustbox}{width=0.45\textwidth}
\input{tikz/PBA_trace}
\end{adjustbox}
\caption{\textbf{Hierarchical PBA} example with four leaf configurations that denotes four states of a certain DSA's firmware.}
\label{fig:PBA_trace}
\end{figure}
    
However, as the potential configurations can be extensive, we propose an enhancement to the existing PBA scheme that we call \emph{hierarchical-PBA} that executes during the DSA boot up phase. All \fdu{}s on the same device can inherit the device properties alongside \fdu{}-specific configurations that can be injected by the user before the enclave execution begins.
Hierarchical-PBA maintains one PCR bank that reflects a chain of properties that are initiated from the user-generated attestation challenge. 
Figure~\ref{fig:PBA_trace} shows an example where a DSA can have four different configuration represented by four leaf nodes. Every configuration change (node in the tree) triggers a \texttt{TPM2\_pcrExtend} ($PCR_{new}\leftarrow SHA(PCR_{old} \| value)$) that produces an unique PCR value. 
The SM extends PCR bank every time there is a new security policy enforcement on the code. 
The user can maintain a list of ``whitelisted' configuration values provided by the hardware vendor. 
Therefore, given a leaf node value, the user can determine if a certain node has been initialized from a proper state.

One concrete example of an attestation report from an AI accelerator is given in~\Cref{fig:acc_report}, and corresponding security properties (part of the enclave manifest) of the same accelerator is depicted in~\Cref{fig:accManifest}. 
This report is sent back to the CPU-SM and then the CPU-SM sends the consolidated attestation report back to the remote verifier.

\begin{figure}[t]
\begin{lstlisting}[language=json,firstnumber=1]
{"device" : "Accelarator1", "manufacturer" : "Manuf1",
"firmwareVersion" : "X.YZ",
"deviceType" : "AI-Acc", "memoryRegion" : "256M"
"securityCapability" : {
  "debug" : FALSE,
  "coreIsolation" : TRUE, 
  "privateCache" : TRUE,
  "sharedCache" : FALSE, 
  "sharedCacheIsolation" : FALSE,
  "sharedMemoryIsolation" : TRUE,
  "rowHammerMitigation" : FALSE
   }, 
  "Signature":"0x5005feb12c..."}
\end{lstlisting}
\caption{Example security properties of an accelerator, part of an enclave manifest.}
\label{fig:accManifest}
\end{figure}

\begin{figure}[t]
\begin{lstlisting}[language=json,firstnumber=1, escapeinside = ||]
{"DeviceID" : "0x71f70e16a9ff9ca4ba14...",
"PublicKey" : "0xdcc026850848929d8b3...",
"Manufacturer" : "Manuf1",
"FirmwareVersion" : "X.YZ",
"DeviceType" : "type1",
|\textcolor{codegreen}{\# Accelerate properties from the manifest ((refer to the example in \Cref{fig:accManifest})}|%
"SecurityProperties" : {...},
|\textcolor{codegreen}{\# Platform configuration register values}|
"Configuration" : {
    "PCR0" : "0xdcc0",
    "PCR1" : "0xaeff",
    ...
    "PCR23" : "0x1001"
} 
"PolicyQuote" : "0xd0299944598e0f...",
|\textcolor{codegreen}{\# FDU properties)}|
"FDU" : {
    "FDU_Support" : TRUE,
    "Numbers" : 5,
    "Config" : [
       {"BAR" : "0xffff3225",
        "streamID" : "0xc1fc8cd84",
        "core_config" : 2,
        "memory" : "128M"},
        {...} 
    ]    
},
"FirmwareQuote":"0xdcc026850848929d8b3...",
"Certificate":"0x5005feb12ca9842eb60c8..."}
\end{lstlisting}
\caption{Example attestation result of an accelerator.}
\label{fig:acc_report}
\end{figure}

\myparagraph{Key Distribution} As part of the attestation process, the \user receives per-FDU public keys that it can use to setup secure channels. 
Usually, along with the run command to start the computation, \user needs to send confidential data for the computation as well. 
For this, it first generates per-FDU public-private key pairs. 
Then, it uses the per-FDU public keys to generate a per-FDU shared key (e.g., Diffie-Hellman ). This key is used to establish a secure channel with every FDU. 
Further, the \user has to provision a shared communication key for all the FDUs. This communication key is used to send and receive data and commands securely over the cloud infrastructure between the FDUs. 
Finally, the \user encrypts the data, run command, shared communication key using the per-FDU shared secret. To this it adds the per-FDU public key it generated and \manifestid to the message and integrity protects it. It then sends the message to the management plane.
The management plane forwards the message to the respective machines. The hypervisors on the individual machines relay the message to the enclaves on the trusted FDUs. The enclave first computes the shared secret using its own private key, provisioned by the SM at startup and the public key of \user in the message. Using this it decrypts the data and the shared key for communication with other FDUs.

\myparagraph{Revocation} We assume either the CSP or the user maintains a revocation list of host, devices and their firmware versions. Such a list can be polled from the CPU or device manufacturer signed by their root key.

\subsection{Case for not using software-based attestation} 
\label{sec:appendix:sw-attestation}

In our proposal, we decided not to leverage any existing software-based attestation~\cite{seshadri2004swatt,li2010sbap,seshadri2008sake,yang2007distributed} for both TEE and non-TEE nodes due to following reasons:
\begin{mylist}
    \item Most of the existing software-based attestation mechanism assumes that the underlying platform does not support hardware TEE primitives. Such an assumption is valid for low-power platforms such as wireless sensor networks or embedded systems. However, in data centers, powerful nodes have massive compute capabilities and capable of supporting TEE primitives.
    
    \item Existing software-based attestations are limited to simpler embedded firmware, whereas we assume a more complex, powerful device firmware that is capable of executing complex workloads. Such a device requires a full-fledged measurement-based attestation.
    
    \item Existing research~\cite{castelluccia2009difficulty} shows software-based attestation are vulnerable to rootkit-based and compression attacks.
\end{mylist}

Given the above-mentioned reasons, we conclude that the software-based attestation of the node is unsuitable for data center/cloud scenario.

\subsection{Existing Device TEEs}
\label{sec:appendix:existing_tees}

\subsubsection{GPUs} 
\label{sec:appendix:existing_tees:gpu}

Graviton~\cite{volos2018graviton} instantiates a TEE for NVIDIA GPUs to enable isolated execution of kernels by using a trusted security monitor (Sm) that processes all commands to the GPU. Graviton partitions the regions that can be mapped by the host (for MMIO) into unprotected, protected and hidden regions. To create these regions, it augments the PCIe controller to perform range checks that are initialised by the SM at start up. The SM ensures that memory allocated to a kernel is not accessible to other non-trusted entities. To guarantee memory isolation, the SM tags pages with ids of the kernels they are allocated to and closely tracks all page mappings. All allocation and de-allocations are handled by the SM and the untrusted driver has no direct access to GPU memory. 
However, the GPU runtime is trusted and the runtime API is augmented to enable secure operations on the GPU. 

Applications have be modified to use the secure API with the runtime. 
To start a \job, the driver initialises a secure context using the CH\_CREATE call to the SM. This creates a protected command channel inaccessible to the driver. 
This call is also used to perform attestation, using CH\_MEASURE and setup a shared secret for channel encryption (CEK) between the host application and the GPU. 
The driver uses CH\_PDE to allocate memory for the secure context on the GPU. 
To load a kernel and start computation, the host application uses the secure runtime API. The runtime encrypts the kernel code and data and initiates a DMA to transfer the encrypted data to the region it was allocated on the GPU. 
The runtime then uses an Authenticated decryption kernel on the GPU to decrypt the data and code and store it in protected private memory. 

Graviton performs encrypted DMA to enclave private memory. The driver cannot perform MMIO to the command channels as it does not have access to the CEK. 
Therefore, Graviton forces the runtime to write to the command channel using DMA. The runtime copies encrypted commands to the GPU, decrypts it using the AuthDec kernel and then places it in the command channel. 
The driver can still perform MMIO to GPU's non-protected memory. The SM uses the unprotected space to notify the device driver about errors, synchronization commands and interrupts. 

\subsubsection{FPGA} 
\label{sec:appendix:existing_tees:fpga}

ShEF~\cite{zhao2021shef} creates a TEE for FPGAs. Unlike computation on CPUs, FPGAs are programmed using confidential bitstreams (binaries).
ShEF ensures confidentiality and integrity of both the bitstream and data. FPGAs deployed in the cloud usually contain a CSP controlled Shell, that provides the API for user bitstreams to communicate with the host. 

Under the untrusted CSP threat model, ShEF isolates the user program from the untrusted Shell. ShEF partitions the hardware into two \textit{isolated} regions: one for the Shell, and another for the trusted user bitstream.
For secure boot, ShEF requires a hardware root-of-trust on the FPGA. It also requires a hardware SM that uses the RoT to boot the FPGA correctly and generate attestation reports. 
The attestation reports are then sent via the Shell to the user who verifies it and establishes a secure channel with the SM. The user then sends the confidential bitstream over the secure channel to the SM. 
The SM loads the user's bitstream in the region it created and loads the communication key into the Shield. The user can now directly communicate with the bitstream and send data using DMA over a secure channel using the communication key. 

The Shield ensures that all data leaving the user's bitstream region is encrypted with the communication key. Similarly, the Shield provides a separate interface for MMIO registers. The user can directly write encrypted commands or data to the MMIO registers that the Shield decrypts before the write completes. ShEF requires the developer to add a trusted Shield to their application before the bitstream is created. This trusted Shield performs the encryption/decryption for all data that leaves or enters the user bitstream. 

\subsection{Architecture of Ascend AI Accelerator}
\label{sec:appendix:acc}

 \begin{figure}[t]
   \centering
     \includegraphics[trim={0 13.5cm 17.8cm 0},clip,width=0.9\linewidth]{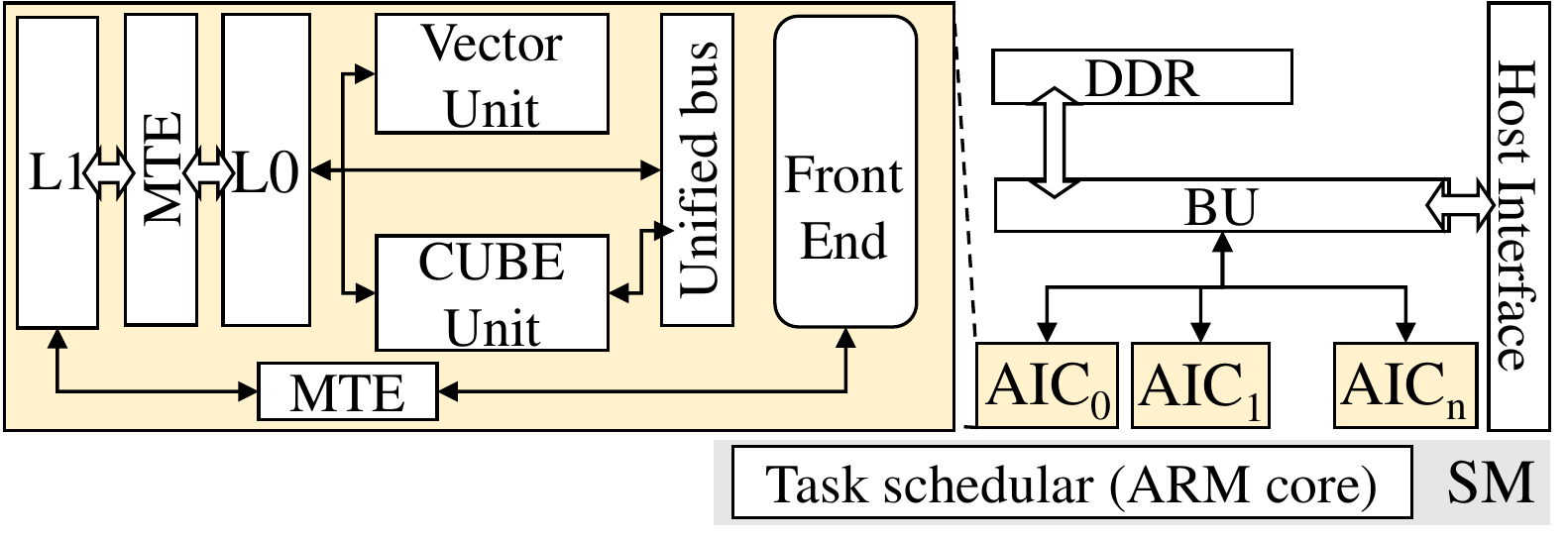}
     \caption{\textbf{AI Accelerator Case study}. An overview of the AI accelerator based on DaVinci~\cite{liao2019davinci,liao2021ascend} architecture that shows the core, memory subsystems , broadcast unit (BU), task scheduler, and the host interface.}
     \label{fig:ai_acc_architecture}
 \end{figure}

\Cref{fig:ai_acc_architecture} shows a simplified base model of AI accelerator with $n$ cores based on the DaVinci architecture~\cite{liao2019davinci}. Each core is capable of running CNN layers and exposes a custom ISA. The \textit{Cube Unit} is responsible for matrix multiplication and the \textit{Vector unit} is responsible for vector operation. The cube and vector units are connected to the global memory (DDR) via a unified buffer (UB). The units fetch data through two layers of software-managed caches $L0$ and $L1$. There are two data mover units (known as memory transaction engine or \textit{MTE}). One of the MTEs moves data between L1 and L0 and another between L1 and the global DDR through the core front-end. The MTEs are event-driven, i.e., MTEs move the data from the global memory to the computing units when they need the data and writes it back to the global memory when computation is done. The precise sequence of such operations (execute, move data) is determined by an event scheduler on the accelerator SoC. The scheduler is a single-stage in-order ARM core that receives the compiled kernel instructions from the host and assigns them to the cores. The cores rely on an in-order scalar front-end to offload instructions to the MTEs, the Vector Unit, and the Cube Unit. The cores are connected to a broadcast unit (BU). The BU collects memory requests from the cores and responses from the memory subsystem and routes them to the correct endpoint. A detailed design of the accelerator is described in~\cite{andri2022winograd}.

\begin{table}[t]
\caption{SSD configurations from SimpleSSD~\cite{simplessd}}
\spacesave
\begin{adjustbox}{width=\columnwidth,center}
\begin{tabular}{llll}\toprule
SSD                             & ARM core Freq & Cache size & NVM controller BW\\\midrule   
Intel 750 400GB                 & 400 MHz   & 800MB   & 3.2 GB/s\\
Samsung Z-SSD prototype 800GB   & 800 MHz   & 1GB     & 4 GB/s\\
Samsung 983 DCT 1.92TB          & 800 MHz   & 512 MB & 696 MB/s\\
Samsung 850 PRO 256GB           & 400 MHz   & 384 MB & 4GB/s \\
Intel 535 240GB                 & 300 MHz   & 256 MB & 2GB/s\\
\bottomrule
\end{tabular}
\end{adjustbox}
\label{tab:ssd_configs}
 \end{table}

\subsection{Architecture of Modern SSDs}
\label{sec:appendix:ssd_arch}

 SSDs typically contain host interface layer (HIL) that implements a interface such as PCIe subsystem to talk to the CPU, a SSD controller, and a CPU core (typically an ARM-M profile core) that runs the firmware of the SSD controller. The SSD controller consists of an internal cache layer (ICL) and page-level flash translation layer (FTL). ICL manages the DRAM cache speeds up the IO transaction by caching recently used pages. ICL implements cache replacement policies. FTL implements the flash translation, i.e., converting logical page numbers to physical page numbers (both are 4K pages), garbage collection, SSD trimming, die wear level management, and over-provisioning management.

%% file: tikz/PBA_trace.tex
    \begin{tikzpicture}[root/.style={rectangle,draw,fill=purple!25, rounded corners, align=center},
    first/.style={rectangle, draw,fill=orange!40, rounded corners, align=center},
    second/.style={rectangle, draw, rounded corners, align=center, fill=black!5},
    edge from parent/.style={draw,-latex},
    sibling distance=10em]]

    \node[root](root) {ROOT: \texttt{0x235aa...}}
    child { node[first, yshift=1.8em, xshift=-3em] (de) {Debug disabled: \texttt{0x566f..}} 
      child { node[second, yshift=1.6em] (c1) {Core sep\\\texttt{0xab20..}}}
      child { node[second, yshift=1.6em, xshift=-3em] (m1) {Mem sep\\\texttt{0x85ff..}} }}
    child { node[first, yshift=1.8em,xshift=1em] (dd) {Debug enabled : \texttt{0xd225f....}}
      child { node[second, yshift=1.6em,xshift=3em] (c2) {Core sep\\\texttt{0x5af3..}}}
      child { node[second, yshift=1.6em] (m3) {Mem sep\\\texttt{0x322a..}} } };
      
    \node[right=10em of root, rotate=-90](rt) {PCR[23] Extend};
    \node[left=1.6em of root](nonce) {Random challenge};
    \node[left=1.8em of root](p) {};
    \node[right=1.5em of p](q) {};
    \draw [-latex](p) -- (q);
    
    \node[right=9em of root](a) {};
    \node[below=6.5em of a](b) {};
    \draw [-latex](a) -- (b);
      
    \end{tikzpicture}

  